\documentclass[twocolumn,secnumarabic,amssymb, nobibnotes, aps, prr]{revtex4-2}
\usepackage{braket}
\usepackage{physics}
\usepackage{mathtools}
\usepackage{amsfonts}
\usepackage{amsmath}
\usepackage{tabularx} % in preamble
\usepackage{graphicx}% Include figure files
\usepackage{dcolumn}% Align table columns on decimal point
\usepackage{bm}% bold math
\usepackage{diagbox}

\usepackage{float}
\usepackage{xcolor}
\usepackage{soul} % For strikeout text.
\usepackage{placeins} %To force the thins to be after other things. \FloatBarrier

\newsavebox{\mstrut}

\newcommand{\scom}[1]{%
    \sbox{\mstrut}{\(#1\)}%
    \mathinner{\left[\mkern-2mu\left[{#1}\right]\kern-0.1\ht\mstrut\right]}%
}

\begin{document}

\title{Algebraic structure of Fock-state lattices}%

\author{Piergiorgio Ferraro}
\email[]{piergiorgio.ferraro@edu.unito.it}
\affiliation{Department of Physics, University of Turin, via P. Giuria 1, 10125 Turin, Italy}
\author{Caio B. Naves}%
\email[]{caio.naves@fysik.su.se}
\affiliation{Department of Physics,
Stockholm University, AlbaNova University Center, 106 91 Stockholm,
Sweden}
\author{Jonas Larson}%
\email[]{jolarson@fysik.su.se}
\affiliation{Department of Physics, Stockholm University, AlbaNova University Center, 106 91 Stockholm, Sweden}
\begin{abstract}
We analyze Fock-state lattices (FSLs) from an algebraic viewpoint. Starting from a Lie algebra, we associate a FSL constructed from the action of its generators: diagonal (Cartan) generators define the lattice sites, while off-diagonal (root) generators determine the lattice bonds. This construction reveals that identifying an underlying algebraic structure provides direct physical insight into FSLs, including their dimensionality, connectivity, symmetry constraints, and possible transport and revival phenomena. By examining several common Lie algebras, we identify not only their associated FSLs but also the corresponding Lie phase spaces, thereby establishing a systematic connection between FSL dynamics and phase-space geometry. In many cases, both the phase space and the FSL exhibit nontrivial curvature, opening possibilities for exploring quantum dynamics in curved synthetic spaces. We further address whether every integrable Hamiltonian admits an underlying Lie algebra that reproduces the same FSL structure. We show that this is not generally the case, particularly for Hamiltonians that are nonlinear in the generators, and that for systems combining different types of degrees of freedom the appropriate underlying structure may instead be a Lie superalgebra.
\end{abstract}

%\pacs{45.50.Pq, 03.65.Vf, 31.50Gh}
%\date{}%
\maketitle
%\tableofcontents

%%%%%%%%%%%%%%%%%%%%%%%%%%%%%%%%%%%%%%%%%%%%%%%%%%%%%%%%%%%%%%%%%%%%%%%%%%

\section{Introduction} 
In the past few decades, quantum simulators have become ubiquitous tools for the simulation of quantum phenomena~\cite{georgescu2014quantum,altman2021quantum}. Broadly speaking, a quantum simulator is defined as a controllable quantum-mechanical system that can be used to emulate other quantum systems. Compared to a quantum computer, quantum simulators offer several important practical advantages. First, a quantum simulator can be tailored to a specific problem, thus requiring fewer resources. Second, the high degree of control over the implemented degrees of freedom enables easier preparation of initial states and more efficient readout procedures.

Examples of platforms that exhibit these features include ultracold atoms in optical lattices~\cite{bloch2008many,lewenstein2007ultracold}, trapped ions~\cite{blatt2012quantum,britton2012engineered,schneider2012experimental}, and linear optical systems~\cite{aspuru2012photonic,hartmann2016quantum}. Many of these systems, such as atoms and ions, employ subsystems equipped with highly controllable internal degrees of freedom. Motivated by this, researchers have proposed exploiting these internal degrees of freedom to mimic additional spatial dimensions, leading to the concept of synthetic dimensions~\cite{grass2025colloquium}. This development has opened new avenues to investigate, for example, topological phenomena and other exotic quantum phases of matter~\cite{celi2014synthetic,ozawa2016synthetic,arguello2024synthetic}, particularly in photonic architectures~\cite{ozawa2016synthetic,yu2025comprehensive}, while also extending the capabilities of cold-atom~\cite{celi2014synthetic,price2017synthetic,chen2024strongly,boada2011dirac} and trapped-ion platforms~\cite{bermudez2011synthetic,manovitz2020quantum}.

 The flexibility of synthetic dimensions suggests that this idea can be pushed even further in the construction of quantum simulators. Instead of relying on a real-space lattice, one can devise a lattice that exists purely in Hilbert space. This type of lattice is commonly referred to as a Fock-state lattice (FSL)~\cite{wang2016mesoscopic,saugmann2023fock,larson2021jaynes,yuan_quantum_2024}. 

A FSL is constructed by selecting a quantum system whose internal states—typically expressed in the Fock basis—are interpreted as lattice sites, while the transitions between them are engineered through an appropriate choice of Hamiltonian. As synthetic lattices, FSLs allow one to exploit the available number of states as effective sites. For example, a single quantum harmonic oscillator provides, in principle, an infinite number of such sites. However, an important caveat is that, in many instances, the translational symmetry of the FSL is broken due to the underlying algebra of the corresponding creation and annihilation operators. Fortunately, many properties of interest do not rely on translational symmetry, such as topological characteristics defined by local markers~\cite{cai_topological_2021}, or frustration, which instead relies on macroscopic degeneracy~\cite{moessner2006geometrical}. 

Experimentally, FSLs have been explored in a wide range of contexts, including the realization of topological phases~\cite{cai_topological_2021,deng_observing_2022,wu2023observation,yang2022simulating,sriram2025quantized}, flat bands~\cite{yang2024realization}, synthetic gauge fields and the Aharonov--Bohm effect~\cite{zhang2025synthetic}, and many-body dynamics~\cite{yao2023observation}. Theoretically, FSLs have been connected to Floquet space~\cite{sriram2025quantized,larson2024floquet}, dark states~\cite{zhao2025dark}, the preparation of highly excited Fock states~\cite{li2026scalable}, localization phenomena such as Aharonov--Bohm cages~\cite{ben2025many,jonay2025localized,yao2026non}, and open quantum systems~\cite{naves2025liouville}.

In this work, we present a conceptually different approach to FSLs that does not rely on Hamiltonians, but is instead based on Lie algebras. Specifically, the generators of a Lie algebra are divided into Cartan generators -- those forming a set of mutually commuting and diagonalizable operators -- and root generators. From the diagonal Cartan generators, one identifies a weight lattice, which establishes a correspondence to Fock states. The root generators, being non-diagonal in the Fock basis, induce transitions between the sites of the weight lattice. In this manner, a FSL can be constructed directly from the Lie algebra.  

This approach has the advantage that algebraic techniques can be applied to FSLs. For example, one can associate a phase space and generalized coherent states with the FSL, providing alternative descriptions of its structure. By considering commonly studied Lie algebras, we identify the corresponding phase spaces and FSLs, and explore their dynamics after assigning a Hamiltonian. For a general algebra, the corresponding phase space is not flat, reflecting the lack of translational invariance of the FSL. This demonstrates that FSLs can simulate dynamics in curved spaces. While one-dimensional lattices can always be embedded in flat space, higher-dimensional lattices—such as those arising from $\mathfrak{su}(3)$ and $\mathfrak{so}(5)$—cannot be embedded in this way. 

The Hamiltonian and Lie-algebra frameworks are related: given a Lie algebra, one can construct a Hamiltonian as a linear combination of its generators. However, this relation is not invertible: for a general Hamiltonian, the existence of a corresponding Lie algebra is not guaranteed. For example, obtaining a finite FSL requires that the Hamiltonian be integrable, but integrability alone does not ensure a closed Lie algebra. This consideration holds for Hamiltonians involving the same type of degrees of freedom. For integrable models involving multiple types of degrees of freedom, a conventional Lie algebra may not suffice, though in some cases a correspondence between Hamiltonians and algebras can be restored by considering Lie superalgebras.

The outline of the paper is as follows. In the next section, we review the Hamiltonian formulation of Fock-state lattices and introduce the key concepts of Lie algebras in Subsec.~\ref{ssec:lie}. In Sec.~\ref{sec:lie}, we establish the general correspondence between Lie algebras and Fock-state lattices, followed by several examples ranging from algebras with well-known FSLs, such as the Euclidean and Heisenberg--Weyl algebras, to others that are less familiar from the Hamiltonian perspective. We explore the relationship between the Hamiltonian and Lie-algebra approaches in Sec.~\ref{sec:discussion}. Finally, in Sec.~\ref{sec:conclusion}, we summarize our findings and discuss future directions.

\section{Background: Fock-state lattices and Lie algebras}
\label{sec:background}
%%%%%%%%%%%%%%%%%%%%%%%%%%%%%%%%%%%%%%%%%%%%%%%%%%%%%%%%%%%%%%%%%%%%%%%%%%%%%%%%%%%%%%%%%%%%%%%%%%%%%%%%%%%%%%%%%

\subsection{Fock-state lattices: Hamiltonian description}
\label{ssec:fsl}

Fock-state lattices (FSLs) provide an example in which physics is realized in synthetic dimensions rather than in real space. In an FSL, lattice sites correspond to physical states of a quantum system instead of spatial positions. The term ``Fock state'' should therefore not be interpreted literally; rather, it refers more generally to number-like basis states relevant to the system under consideration. Examples include bosonic Fock states $|n\rangle$ ($n \in \mathbb{Z}_{\ge 0}$), fermionic Fock states $|n_1,n_2,\dots\rangle$ ($n_i=0,1$), spin or angular-momentum states $|S,m\rangle$ ($-S\le m\le S$), or site-localized lattice states $|j\rangle$ ($j\in\mathbb{Z}$).

Any Hamiltonian $\hat{H}$ can be represented as a matrix $\bar{H}$ in this basis, with elements
$H_{\{nm\}}=\langle\{n\}|\hat{H}|\{m\}\rangle$,
where $|\{n\}\rangle$ denotes a basis state labeled by the multi-index $\{n\}$. For a single degree of freedom, $\{n\}$ reduces to a single integer, whereas for multiple degrees of freedom it collects all relevant quantum numbers. Each basis state $|\{n\}\rangle$ is associated with a vertex (lattice site) with onsite energy
\begin{equation}
    \varepsilon_{\{n\}}\equiv\langle\{n\}|\hat{H}|\{n\}\rangle.
\end{equation}
Edges in the lattice are determined by allowed transitions,
\begin{equation}
    J_{\{nm\}}\equiv\langle\{n\}|\hat{H}|\{m\}\rangle\neq 0,
\end{equation}
where $|J_{\{nm\}}|$ plays the role of a tunneling amplitude. By definition, $\varepsilon_{\{n\}}=J_{\{nn\}}$. The Hamiltonian $\bar{H}$, together with the chosen basis, therefore defines a graph that we refer to as a \emph{Fock-state lattice}. Explicitly,
\begin{equation}
    \hat{H}
    =\sum_{\{n\}}\varepsilon_{\{n\}}|\{n\}\rangle\!\langle\{n\}|
    +\sum_{\{n\}\neq\{m\}}J_{\{nm\}}|\{n\}\rangle\!\langle\{m\}|.
\end{equation}

Writing the state as
\begin{equation}
    |\Psi(t)\rangle=\sum_{\{n\}} c_{\{n\}}(t)|\{n\}\rangle,
\end{equation}
and collecting the coefficients into the vector $\bar{C}(t)$, the Schrödinger equation becomes
\begin{equation}
    i\frac{d\bar{C}(t)}{dt} = \bar{H}\,\bar{C}(t),
\end{equation}
with site populations
\begin{equation}
    P(\{n\},t) = |c_{\{n\}}(t)|^2.
\end{equation}
Within the FSL framework, the physical problem is thus mapped onto that of a single particle hopping on a lattice, whose dimensionality, connectivity, and hopping amplitudes are determined by the underlying physical model. For many-body systems with many degrees of freedom, the resulting lattice is typically complex, high dimensional, and exhibits long-range hopping~\cite{basko2006metal}. The power of FSLs is therefore most evident in systems with few, well-isolated, and controllable degrees of freedom, as commonly realized in quantum-optical settings such as trapped ions, cavity or circuit QED, and nonlinear optics~\cite{saugmann2023fock,yuan2024quantum}.

To illustrate these ideas, we consider a simple yet nontrivial example: two degenerate bosonic modes described by the Hamiltonian
\begin{equation}\label{su2ham}
    \hat{H}_\mathrm{2m}=J_0\left(\hat{a}^\dagger\hat{b}+\hat{b}^\dagger\hat{a}\right).
\end{equation}
This model can be realized, for instance, in a bimodal cavity containing a two-level system dispersively coupled to two degenerate cavity modes~\cite{larson2021jaynes}. Defining the detuning $\Delta=\Omega-\omega$, where $\Omega$ is the atomic transition frequency and $\omega$ the cavity frequency, and assuming $|\Delta|\gg g\sqrt{\bar{n}}$ (with $\bar{n}$ the average photon number), second-order perturbation theory yields an effective coupling of the above form with $J_0=g^2/\Delta$.

The total excitation number
\begin{equation}
    \hat{N}=\hat{n}_a+\hat{n}_b,
\end{equation}
with $\hat{n}_a=\hat{a}^\dagger\hat{a}$ and $\hat{n}_b=\hat{b}^\dagger\hat{b}$, is conserved. The Fock states $|j\rangle\equiv|n_a=j,n_b=N-j\rangle$, with $j=0,1,\dots,N$, can therefore be labeled by a single integer and grouped into sectors of fixed $N$. As a result, the FSL consists of a set of one-dimensional chains, one for each excitation sector. Each chain is finite, and the tunneling amplitudes
\begin{equation}
\begin{aligned}
    J_{j,j+1}&=\langle j+1|\hat{H}_{2m}|j\rangle
    =J_0\sqrt{(j+1)(N-j)},\\
    J_{j,j-1}&=\langle j-1|\hat{H}_{2m}|j\rangle
    =J_0\sqrt{j(N-j+1)}
\end{aligned}
\end{equation}
depend explicitly on the site index $j$. Tunneling beyond nearest neighbors vanishes, so that the lattice is of tight-binding type.

Further insight is obtained using the Schwinger-boson mapping~\cite{sakurai2020modern,auerbach2012interacting}
\begin{equation}
    \hat{S}_z = \frac{1}{2}\left(\hat{a}^\dagger\hat{a}-\hat{b}^\dagger\hat{b}\right),\quad
    \hat{S}^+ = \hat{a}^\dagger\hat{b},\quad
    \hat{S}^- = \hat{b}^\dagger\hat{a},
\end{equation}
which casts the Hamiltonian into the simple form
\begin{equation}
\hat{H}_{2m}=2J_0\,\hat{S}_x=J_0\left(\hat{S}^++\hat{S}^-\right).
\end{equation}
Its eigenstates are $|S,m\rangle_x$ with $-S\le m\le S$, where $S=N/2$. The spectrum is finite and equally spaced, $\varepsilon_m=2J_0 m$, implying perfect revivals of any initial state with revival time $T_{\mathrm{rev}}=\pi/J_0$.

It is straightforward to generalize this model to three modes,
\begin{equation}\label{su3ham}
    \hat{H}_{3m}
    =J_0\left(\hat{a}^\dagger\hat{b}
    +\hat{b}^\dagger\hat{c}
    +\hat{a}^\dagger\hat{c}
    +\text{H.c.}\right).
\end{equation}
The total excitation number
$\hat{N}=\hat{n}_a+\hat{n}_b+\hat{n}_c$
is again conserved. The corresponding FSL is a finite, tight-binding two-dimensional triangular lattice~\cite{saugmann2023fock}. The Schwinger-boson construction generalizes naturally to multimode systems~\cite{anishetty2009irreducible}. For three modes, the mapping can be expressed as
\begin{equation}\label{su3boson}
    \begin{aligned}
         \hat{H}_1 &= \frac{1}{2}\left(\hat{a}^\dagger\hat{a}-\hat{b}^\dagger\hat{b}\right),\\
         \hat{H}_2 &= \frac{1}{2\sqrt{3}}\left(\hat{a}^\dagger\hat{a}+\hat{b}^\dagger\hat{b}-2\hat{c}^\dagger\hat{c}\right),\\
         \hat{I}_{+} &= \hat{a}^\dagger\hat{b},\quad
         \hat{U}_{+} = \hat{b}^\dagger\hat{c},\quad
         \hat{V}_{+} = \hat{a}^\dagger\hat{c},
    \end{aligned}
\end{equation}
with $\hat{I}_{-}$, $\hat{U}_{-}$, and $\hat{V}_{-}$ given by the Hermitian conjugates of the corresponding raising operators. The notation follows standard conventions. Expressed in terms of these operators, the Hamiltonian takes the form
\begin{equation}
    \hat{H}_{3m}=J_0\left(\hat{I}_++\hat{I}_-+\hat{U}_++\hat{U}_-+\hat{V}_++\hat{V}_-\right).
\end{equation}

In the two-mode model, the introduction of a complex phase factor,
$\hat{S}_\pm \rightarrow e^{\pm i\phi}\hat{S}_\pm$, has no physical effect, since it can be removed by an appropriate unitary rotation generated by $\hat{S}_z$. This reflects the fact that the phase corresponds to a global gauge choice.

In contrast, for the three-mode model, introducing a phase factor in any one of the operators $\hat{I}_\pm$, $\hat{U}_\pm$, or $\hat{V}_\pm$ cannot, in general, be eliminated by a unitary transformation. From the FSL perspective, such a phase renders the tunneling amplitudes complex and leads to nontrivial phases accumulated around closed loops of the lattice. This is directly analogous to the Peierls substitution for a charged particle hopping on a lattice in the presence of a perpendicular magnetic field, with the phase factors corresponding to an effective magnetic flux piercing the plaquettes of the FSL.

%%%%%%%%%%%%%%%%%%%%%%%%%%%%%%%%%%%%%%%%%%%%%%%%%%%%%%%%%%%%%%%%%%%%%%%%%%%%%%%%%%%%%

\subsection{General concepts of Lie algebras}
\label{ssec:lie}

To establish the connection between Lie algebras and FSLs in the following section, we briefly review the necessary elements of Lie-algebra theory.

Let $\hat{X}_a$ denote the generators of a Lie algebra $\mathfrak{g}$, obeying the commutation relations~\cite{georgi2000lie,gilmore2006lie}
\begin{equation}\label{liecom}
    [\hat{X}_a,\hat{X}_b] = i{f_{ab}}^{c}\,\hat{X}_c,
\end{equation}
where ${f_{ab}}^{c}$ are the \emph{structure constants}. Exponentiation of the generators $\hat{X}_a$, or of their linear combinations, yields elements of the associated Lie group $G$.

The \emph{rank} $r$ of the algebra is defined as the number of mutually commuting generators that can be simultaneously diagonalized. These generators form the \emph{Cartan subalgebra} and are denoted $\hat{C}_a$, $a=1,\dots,r$. The remaining generators are \emph{root generators} $\hat{E}_{\alpha}$, labeled by roots $\alpha$. The roots occur in pairs $\pm\alpha$, and the commutation relations take the general form
\begin{equation}
\left[\hat{E}_{\alpha_i}, \hat{E}_{\alpha_j}\right] =
\begin{cases}
\displaystyle \sum_{a} (\alpha_i)_a \, \hat{C}_a,
& \text{if } \alpha_i = -\alpha_j, \\[6pt]
\displaystyle N_{\alpha_i,\alpha_j}\, \hat{E}_{\alpha_i+\alpha_j},
& \text{if } \alpha_i+\alpha_j \text{ is a root}, \\[6pt]
0, & \text{otherwise},
\end{cases}
\end{equation}
where $(\alpha_i)_a$ denotes the component of the root vector $\alpha_i$ along the $a$th Cartan direction. The total number of root pairs is therefore $(\mathrm{dim}\,\mathfrak{g}-r)/2$. In a \textit{semisimple} Lie algebra, the Cartan generators $\hat{C}_a$ are the only mutually commuting elements. All other generators have nontrivial commutation relations that connect them to the rest of the algebra, so that there are no Abelian ideals. An \textit{ideal} is a subspace $\mathfrak{I}$ of the algebra such that any commutators involving an element in $\mathfrak{I}$ will also be in $\mathfrak{I}$. The simplest example would be if the identity $\hat{\mathbb{I}}$ is part of the algebra, then it forms an ideal. Any generator commutes with $\hat{\mathbb{I}}$ such that it is automatically in $\mathfrak{I}=\mathrm{span}\{\hat{\mathbb{I}}\}$. An element commuting with every other element of the algebra is also called a \textit{central element}. 

Given a Lie group $G$, one can associate a Lie-algebra phase space (LPS) as follows. Consider a reference state $|\psi_0\rangle$ that is annihilated by all root generators corresponding to positive roots,
\begin{equation}
    \hat{E}_{\alpha_i} |\psi_0\rangle = 0, \qquad \forall\, \alpha_i>0.
\end{equation}
We define the displacement operators
\begin{equation}\label{gendis}
    \hat{D}_{\alpha_i}(\beta)
    = \exp\!\left(\beta \hat{E}_{\alpha_i}
    - \beta^* \hat{E}_{-\alpha_i}\right),
    \qquad \alpha_i>0,
\end{equation}
and the corresponding family of states
\begin{equation}
    |\psi_{\alpha_i}(\beta)\rangle
    = \hat{D}_{\alpha_i}(\beta)\,|\psi_0\rangle.
\end{equation}
The set of all such states forms a manifold, referred to as the Lie-algebra phase space, in which each point corresponds to a unique quantum state.

When there is only a single pair of root generators (so that the root label can be omitted), the states $|\psi(\beta)\rangle$ are known as \emph{Perelomov coherent states} (PCSs)~\cite{perelomov1977generalized,Perelomov1986}. Well-known examples include bosonic coherent states associated with the Heisenberg--Weyl algebra and spin coherent states associated with $\mathfrak{su}(2)$. In general, PCSs are not eigenstates of the root generator $\hat{E}_{\alpha}$; the Heisenberg--Weyl algebra, for which Perelomov and Glauber coherent states coincide, is a notable exception.

For Lie algebras with more than one pair of root generators (and hence multiple Cartan generators), one constructs a generalized displacement operator $\hat{\Omega}(\boldsymbol{\beta})$ involving all root generators. Acting with $\hat{\Omega}(\boldsymbol{\beta})$ on the reference state produces a unique generalized coherent state. Thus, for algebras with a single root pair one has $\hat{D}_{\alpha}(\beta)=\hat{\Omega}(\beta)$, whereas for higher-rank algebras this identification no longer holds.

\onecolumngrid

\begin{table}[ht]
\caption{Summary of key Lie-algebra concepts used in constructing Fock-state lattices (FSLs).}
\centering

\setlength{\tabcolsep}{10pt}
\renewcommand{\arraystretch}{2.5}
\small
\begin{tabular}{lll}
\hline\hline
Concept / Symbol & Definition / Role & Physical / FSL Intuition \\
\hline

\shortstack[l]{Cartan generators $\hat{C}_a$\\ \hspace{0cm}} 
& \shortstack[l]{Mutually commuting generators, $a=1,\dots,r$\\ \hspace{0cm}} 
& \shortstack[l]{Label lattice directions \\ (vertex coordinates)} \\

\shortstack[l]{Root generators $\hat{E}_{\pm\alpha}$\\ \hspace{0cm}} 
& \shortstack[l]{Non-diagonal generators connecting \\ eigenstates of Cartan genrators} 
& \shortstack[l]{Induce hopping \\ between lattice sites} \\

\shortstack[l]{Reference state $|\psi_0\rangle$\\ \hspace{0cm}} 
& \shortstack[l]{Annihilated by all positive root generators\\ highest weight} 
& \shortstack[l]{Defines highest site in FSL\\ center site for non-compact algebras} \\

Displacement operator $\hat{D}_{\alpha}(\beta)$ 
& \shortstack[l]{$\exp(\beta \hat{E}_\alpha - \beta^* \hat{E}_{-\alpha})$} 
& \shortstack[l]{Moves along lattice edges; \\ generates coherent states} \\

Rank $r$ 
& Number of Cartan generators 
& \shortstack[l]{Dimension of the FSL \\ underlying lattice directions} \\

\shortstack[l]{Semisimple Lie algebra\\ \hspace{0cm}} 
& \shortstack[l]{A Lie algebra with no ``Abelian'' \\ (commuting) ideal parts} 
& \shortstack[l]{Broken translational \\ invariance of the FSL} \\

Lie-algebra phase space (LPS) 
& Manifold of coherent states $G/H$ 
& \shortstack[l]{Continuous space embedding \\ the discrete FSL} \\

Perelomov coherent states (PCS) 
& \shortstack[l]{$|\psi(\beta)\rangle = \hat{\Omega}_\alpha(\beta)|\psi_0\rangle$} 
& \shortstack[l]{Points in the LPS; smooth \\ interpolation of lattice dynamics} \\

Husimi function $Q(\beta)$ 
& \shortstack[l]{$Q(\beta) \propto \langle \psi(\beta)|\hat{\rho}|\psi(\beta)\rangle$} 
& \shortstack[l]{Quasiprobability distribution on LPS; \\ visualizes population / coherence} \\

\hline\hline
\end{tabular}

\label{tab:summary_lie}
\end{table}
\twocolumngrid

It is important to distinguish the LPS from the conventional phase space (RPS) used in quantum optics, which is defined in terms of canonical quadratures and associated quasiprobability distributions~\cite{scully1997quantum,carmichael2013statistical,schleich2015quantum}. The RPS depends on the specific physical realization and on the choice of canonical variables prior to quantization. By contrast, the LPS is an intrinsic object determined solely by the Lie algebra and its representation, defined as the manifold of generalized coherent states.

To connect the two notions, one may equip the LPS with quasiprobability distributions. For example, the Husimi function for Glauber coherent states $|\alpha\rangle$ is defined as~\cite{scully1997quantum,carmichael2013statistical,schleich2015quantum}
\begin{equation}\label{husimi}
    Q(\alpha)=\frac{1}{\pi}\langle\alpha|\hat{\rho}|\alpha\rangle,
\end{equation}
where the coherent-state parameter is $\alpha=(x+ip)/\sqrt{2}$, with
$x=\sqrt{2}\,\mathrm{Re}(\alpha)$ and $p=\sqrt{2}\,\mathrm{Im}(\alpha)$.
This construction generalizes naturally to Perelomov coherent states,
\begin{equation}\label{genhusim}
    Q(\beta)=w(\beta)\,
    \langle\psi(\beta)|\hat{\rho}|\psi(\beta)\rangle,
\end{equation}
where $w(\beta)$ is a normalization factor, e.g., $w(\beta)=\pi^{-1}$ for the phase space of a particle on a line and $w(\beta)=(2S+1)/(4\pi)$ for spin coherent states. These Husimi functions are defined on the same manifolds as the corresponding LPSs and play a role analogous to quasiprobability distributions in standard phase space~\cite{scully1997quantum,carmichael2013statistical,schleich2015quantum}. The Husimi function comes natural once the PCSs have been introduced, but more generally, a variety of quasiprobability distributions, including generalized Wigner and Glauber--Sudarshan functions, can be defined on LPSs within the Stratonovich--Weyl framework~\cite{agarwal1981relation,brif1998phase,klimov2010general,tilma2016wigner}. 

In Tab.~\ref{tab:summary_lie} we summarize the various quantities introduced in this subsection. In the second column we explain how they are defined, while in the third we give their physical interpretation, and especially their meaning in terms of the FSLs.

%%%%%%%%%%%%%%%%%%%%%%%%%%%%%%%%%%%%%%%%%%%%%%%%%%%%%%%%%%%%%%%%%%%%%%%%%%%%%%%%%%%%%
\section{From Lie algebras to Fock-state lattices}
\label{sec:lie}
The previous section presented the basic ingredients needed to construct a FSL from a Lie algebra. This constitutes a conceptually different approach: rather than starting from a specific physical system and its Hamiltonian, we begin with the algebraic structure itself. As we will discuss in the next section, while a Lie algebra can be used to define a FSL, the reverse is generally not true. That is, starting from a Hamiltonian and its associated FSL does not automatically imply the existence of a corresponding Lie algebra. Even for integrable Hamiltonians, one cannot generally expect an underlying Lie algebra. However, when a Lie algebra does exist, its techniques provide a powerful tool for understanding the FSL physics and for describing the lattice in terms of a phase-space picture. 

The goal of this section is to first demonstrate, in general terms, how a Lie algebra defines a FSL and what the resulting lattice geometry is. We then analyze several common algebras, the types of FSLs they generate, and their potential physical realizations.

\subsection{Fock-state lattices: Lie-algebra description}
\label{ssec:liefsl}
We are now in a position to introduce the FSL associated with a given Lie algebra. Since the Cartan generators $\hat{C}_a$ are diagonal, they define a set of quantum numbers $m_i^a$ that label the Fock states. In the language of Lie algebras, these $m_i^a$ label the sites of the \textit{weight lattice}. For an algebra of rank $r$, the Fock states can thus be written as
\begin{equation}
    |m_i^1, m_i^2, \dots, m_i^r\rangle.
\end{equation}
Consequently, the rank of the Lie algebra determines the dimensionality of the corresponding weight lattice. The FSL is obtained by taking the weight lattice, associating the sites with the corresponding Fock states, and adding the additional structure provided by the root generators. To do this, we note that the root generators $\hat{E}_\alpha$ couple different Fock states, and because they are independent, each site in the FSL has as many non-vanishing tunneling rates as there are root generators. The roots $\alpha$ correspond to these tunneling directions, which in many cases, but not all, coincide with the principle axes of the FSL Taken together, the Cartan and root generators fully determine the lattice structure, with $\hat{C}_a$ specifying the vertices and $\hat{E}_\alpha$ defining the edges. 

The reference state $|\psi_0\rangle$ plays a special role as it is annihilated by all positive root generators (note that the distinction between positive and negative root generators is sometimes arbitrary). Often, the reference state is clear from the physical system: for bosons or fermions, the vacuum serves as $|\psi_0\rangle=|0\rangle$, while for a spin-$S$ it is $|\psi_0\rangle=|S,S\rangle$. For a single particle on an infinite one-dimensional lattice, no reference state exists, which is related to the structure of the LPS.   
 
When describing the FSL in terms of the LPS, it is useful to clarify the action of the displacement operator on PCSs. Analogous to regular phase space in quantum optics (the $x$--$p$ plane for a particle, or the Bloch sphere for a spin-$S$), the reference coherent state is displaced along geodesics, preserving their local shape. In the LPS, successive applications of $\hat{D}_\alpha(\beta)$ also follow geodesics (note that here we keep $alpha$ and $\beta$ fixed during consecutive displacements), and the phase-space distance between states remains constant. PCSs are isotropic minimum-uncertainty states, saturating the Robertson inequality
\begin{equation}
    \Delta \hat{X}_a \, \Delta \hat{X}_b \geq \frac{1}{2} \big| \langle [\hat{X}_a, \hat{X}_b] \rangle \big|,
\end{equation}
and are locally Gaussian when expressed in local canonical (\textit{Darboux}) coordinates on the Lie-algebra phase space. Localization in the LPS does \textit{not} automatically imply localization in the FSL, though in many examples it does. Similarly, semisimple Lie algebras (and some others) admit bosonic representations, allowing a description in multi-dimensional canonical RPSs; here too, LPS localization does not guarantee RPS localization.

We also note that the Cartan generators $\hat{C}_a$ may also generate displacements along geodesics in certain cases. For example, for a spin-coherent state initialized at the equator of the $\mathfrak{su}(2)$ LPS, a rotation around the $z$-axis, i.e. displaced with the Cartan generator, will displace the state along the equator. However, since they are diagonal in the Fock basis, they do not induce transitions in the FSL, but they add local phase shift akin onsite energy shifts.

One operator not yet mentioned is the Casimir operator, $\hat{\Gamma}$, which can be expressed in terms of the generators $\hat{X}_a$ and is defined by the property that it commutes with all $\hat{X}_a$. Casimir operators are essential for determining the phase-space structure: as `constants of motion', they define invariant surfaces that form the underlying phase space for the coherent states.

Without having to involve FSLs, we can directly see applications of the Lie algebra methods by noting the following. Consider a Hamiltonian $\hat{H}$ that can be expressed as a linear combination of the elements $\hat{X}_a$ of some finite Lie algebra, i.e.
\begin{equation}
    \hat{H}=\sum_{a=1}^{\mathrm{dim}\,\mathfrak{g}}k_a\hat{X}_a.
\end{equation}\label{lieham}
The corresponding Heisenberg equations become
\begin{equation}\label{lieheis}
    \frac{d}{dt}\hat{X}_b=-i\left[\hat{X}_b,\hat{H}\right]=i\sum_{a=1}^{\mathrm{dim}\,\mathfrak{g}}k_a\sum_{c=1}^{\mathrm{dim}\,\mathfrak{g}}f_{ab}^c\hat{X}_c,
\end{equation}
implying that the algebra is close and the Hamiltonian is integrable. A well-known example are the \textit{Bloch equations} for a spin-$S$ particle which can be expressed as $d{\vec{S}}/dt=\vec{\Omega}\times\vec{S}$, where $\hat{H}=\Omega_x\hat{S}_x+\Omega_y\hat{S}_y+\Omega_z\hat{S}_z$ and $\vec{\Omega}=(\Omega_x,\Omega_y,\Omega_z)$. Even if the above result is general, one should be aware of that a closed set of Heisenberg equations might not provide full information of the dynamics, but only about the evolution contained within the span of the Lie algebra. We give an example of this below when discussing the $\mathfrak{su}(2)$ algebra. 

\onecolumngrid

\begin{table}[ht]
\caption{Relation between Lie algebras, their Cartan and root generators, associated phase spaces, and the resulting Fock-state lattices, including semisimplicity.}
\label{tab:algebra_phase_fsl}
\begin{ruledtabular}
\begin{tabular}{llllll}
Algebra & Cartan & Root & Phase space & Fock-state lattice & Semisimple? \\
 & generators & generators &  &  &  \\
\colrule
Euclidean $\mathfrak{e}(2)$
& $\hat{E}_0$
& $\hat{E}^\pm$
& Cylinder
& 1D infinite chain
& Non-semisimple \\
Heisenberg--Weyl $\mathfrak{hw}$
& $\hat{n}$
& $\hat{a},\,\hat{a}^\dagger$
& Plane
& Semi-infinite 1D chain
& Non-semisimple \\
$\mathfrak{su}(2)$
& $\hat{S}_z$
& $\hat{S}^\pm$
& Sphere
& Finite 1D chain
& Semisimple \\
$\mathfrak{su}(3)$
& $\hat{H}_1,\,\hat{H}_2$
& $\hat{I}_{\pm},\,\hat{U}_\pm,\,\hat{V}_\pm$
& Compact 4D manifold
& 2D triangular lattice
& Semisimple \\
$\mathfrak{so}(5)$
& $\hat{H}_1,\,\hat{H}_2$
& $\hat{\Sigma}_{\pm\alpha}$
& Compact 6D manifold
& 2D square
& Semisimple \\
$\mathfrak{su}(1,1)$
& $\hat{K}_0$
& $\hat{K}^\pm$
& Hyperboloid
& Semi-infinite 1D chain
& Non-semisimple \\
\end{tabular}
\end{ruledtabular}
\end{table}
\twocolumngrid

%%%%%%%%%%%%%%%%%%%%%%%%%%%%%%%%%%%%%%%%%%%%%%%%%%%%%%%%%%%%%%%%%%%%%%%%%%%%%%%%%%%%

\subsection{Concrete examples}
\label{ssec:liefslex}
In this subsection we concretize the knowledge of the previous subsection to a set of different models listed in tab.~\ref{tab:algebra_phase_fsl}. By doing so we get an idea how the LPS and FSLs are related, and how to derive the FSL given a Lie algebra. 

%%%%%%%%%%%%%%%%%%%%%%%%%%%%%%%%%%%%%%%%%%%%%%%%%%%%%%%%%%%%%%%%%%%%%%%%%%%%%%%%%%%%%%%%%%%%%%%

\subsubsection{Euclidean algebra}
The Euclidean algebra $\mathfrak{e}(2)$ can be linked to a translationally invariant 1D FSL, i.e., a particle hopping between nearest neighbors along a line. Denoting the lattice site $j$ by $|j\rangle$, the generators can be represented as~\cite{klimov2009group}
\begin{equation}\label{e2gen}
    \hat{E}_0 = \sum_j j\,|j\rangle\!\langle j|, \qquad
    \hat{E}^+ = \sum_j |j+1\rangle\!\langle j|,
\end{equation}
with $\hat{E}^-=\left(\hat{E}^+\right)^\dagger$ and commutation relations
\begin{equation}
    [\hat{E}_0,\hat{E}^\pm] = \pm \hat{E}^\pm, \qquad
    [\hat{E}^-,\hat{E}^+] = 0.
\end{equation}
The standard momentum eigenstates $|k\rangle$ are eigenstates of the root generators,
\begin{equation}\label{e2momentum}
    \hat{E}^\pm |k\rangle = e^{\mp i k} |k\rangle ,
\end{equation}
while the Cartan generator $\hat{E}_0$ acts as the position operator and can be written as $\hat{E}_0 = i\partial_k$. Although the shift operators $\hat{E}^\pm$ commute and have well-defined eigenvalues in momentum space, they are not Cartan generators because they do not label states in the original lattice basis -- they move states between sites rather than measuring a property like $\hat{E}_0$ does.

The action of the displacement operator on a lattice site $|j\rangle$ is
\begin{equation}
\begin{array}{lll}
    \hat{D}(\beta)|j\rangle
    & = & \displaystyle{
    e^{\beta\hat{E}^+ - \beta^*\hat{E}^-}|j\rangle
    = e^{\beta\hat{E}^+} e^{-\beta^*\hat{E}^-}|j\rangle } \\[1.5ex]
    & = & \displaystyle{
    \sum_{m,n=0}^{\infty}
    \frac{\beta^m}{m!} \frac{(-\beta^*)^n}{n!} (\hat{E}^+)^m (\hat{E}^-)^n |j\rangle } \\[1.5ex]
    & = & \displaystyle{
    \sum_{m,n=0}^{\infty}
    \frac{\beta^m (-\beta^*)^n}{m!n!} |j+m-n\rangle } \\[1.5ex]
    & = & \displaystyle{
    \sum_{l=-\infty}^{+\infty}
    J_l(2|\beta|)\,e^{i l \arg\beta}\,|j+l\rangle } ,
\end{array}
\end{equation}
where $l=m-n$ and $J_l(x)$ denotes the Bessel function of the first kind. In the second step, we used the commutativity of the root operators, and in the final step we employed the standard series expansion of the Bessel functions. Since $J_{-l}(x)=(-1)^l J_l(x)$, the probability distribution $|J_l(x)|^2$ is symmetric about $l=0$, and it is predominantly supported within $|l|\lesssim 2|\beta|$, outside of which it decays super-exponentially. 

Strictly speaking, the $\mathfrak{e}(2)$ algebra does not admit a proper reference state $|\psi_0\rangle$ that is annihilated by a root generator. Nevertheless, if we choose $|\psi_0\rangle = |0\rangle$, we obtain
\begin{equation}
    |\beta\rangle
    = \hat{D}(\beta)|0\rangle
    = \sum_{l=-\infty}^{+\infty}
    J_l(2|\beta|)\,e^{i l \arg\beta}\,|l\rangle,
\end{equation}
which may be interpreted as a set of Euclidean PCSs. Using the identity
\begin{equation}
    \int_0^\infty J_m(2r)\,J_n(2r)\,r\,dr
    = \frac{1}{2}\delta_{mn},
\end{equation}
one obtains the resolution of the identity
\begin{equation}
    \int \frac{d^2\beta}{\pi}\,|\beta\rangle\!\langle\beta|
    = \mathbb{I}.
\end{equation}

\begin{figure}[!ht]
    \centering
    \includegraphics[width=0.9\linewidth]{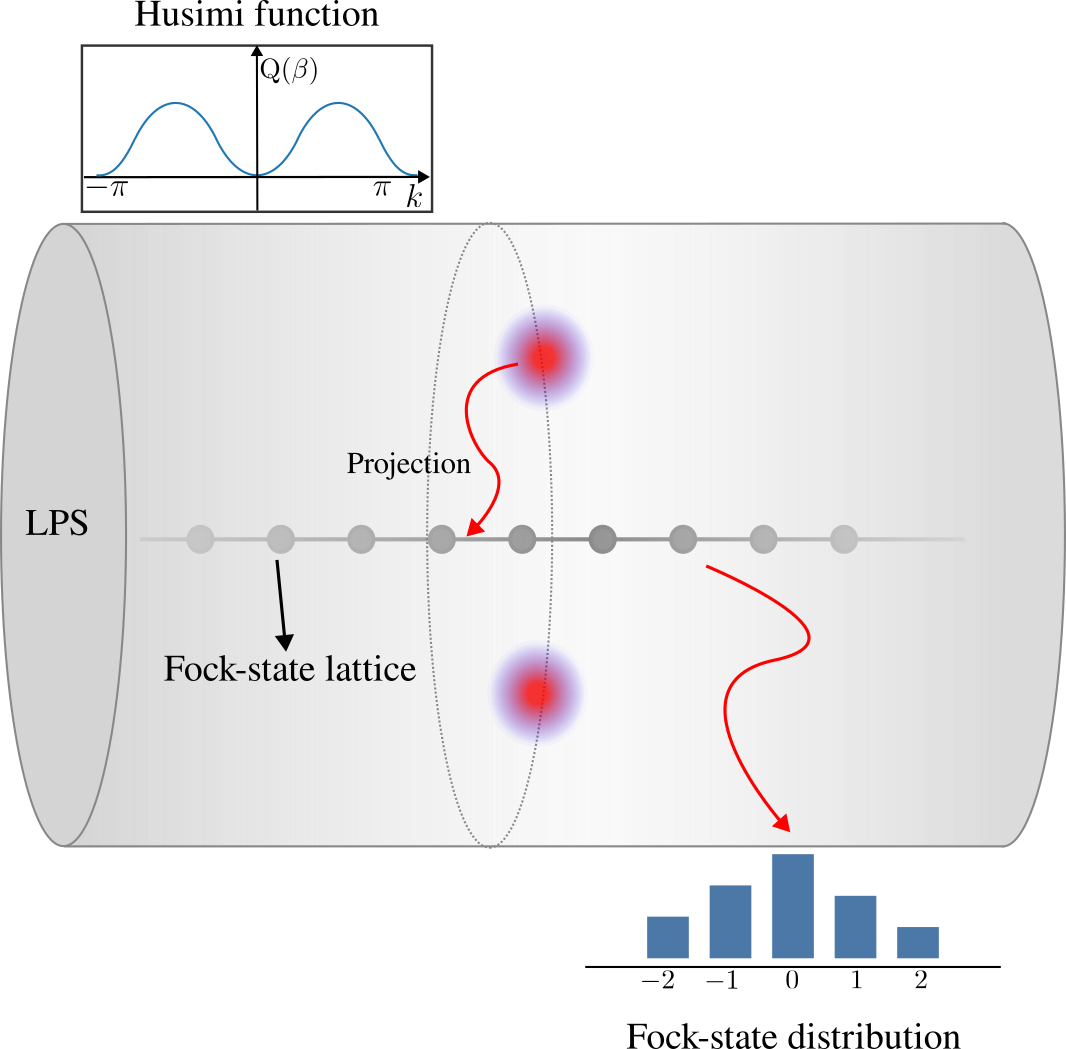}
    \caption{Demonstration of the Lie-algebra phase space (LPS, gray cylinder) and the corresponding Fock-state lattice (FSL, black dots). The Husimi function $Q(\beta)$ is illustrated as the bright structure encircling the LPS cylinder, while the Fock-state distribution $P(j)=\langle j|\hat{\rho}|j\rangle$ is shown as the bar diagram at the bottom. Heuristically, the Fock-state distribution can be viewed as the result of ``projecting'' the Husimi distribution onto the symmetry axis of the cylinder, corresponding to the FSL.}
    \label{fig:e2lps}
\end{figure}

Turning to the LPS of the $\mathfrak{e}(2)$ algebra, the displacement operator $\hat{D}(\beta)$ induces translations along the lattice, with a displacement proportional to $|\beta|$. The Cartan generator, on the other hand, acts diagonally in the Fock basis and generates phase factors, corresponding to translations in a conjugate angular variable. Therefore, the LPS is periodic in one direction and flat in the other, forming a cylindrical geometry. This cylinder reflects the underlying periodicity of the algebra and gives rise to a corresponding Brillouin zone. The FSL can be thought of as lying along the symmetry axis of the cylinder. The probability distribution $P(j)=|\langle j|\psi\rangle|^2$ can then be obtained from a Husimi function $Q(\beta)$ by projecting onto this axis. This is schematically sown in fig.~\ref{fig:e2lps}, where the LPS is the grey cylinder and the sites of the FSL are the black dots along the vertical axis.

To illustrate the connection between the LPS and the FSL, consider the Hamiltonian
\begin{equation}
    \hat{H}_\mathrm{WS} = \omega \hat{E}_0 - J\left(\hat{E}^+ + \hat{E}^-\right).
\end{equation}
This is the famous \textit{Wannier-Stark Hamiltonian}, describing a particle in a 1D tilted tight-binding lattice, where $\omega$ is the tilt strength (corresponding to a constant force) and $J$ is the tunneling rate~\cite{hartmann2004dynamics}. For nonzero $\omega$, the spectrum is the \textit{Wannier-Stark ladder} $\varepsilon_j = \omega j$ for integers $j$. While the spectrum is independent of $J$, the eigenvectors $|\phi_j\rangle$ are exponentially localized around site $j$. The equidistant spectrum implies perfect revivals with period $T_\mathrm{rev} = 2\pi/\omega$. Two characteristic types of evolution are observed: for initial states localized in real space, e.g., $|\psi(0)\rangle = |j\rangle$, the evolved state $|\psi(t)\rangle$ displays a breathing motion; for initial states localized in momentum space, $|\psi(t)\rangle$ exhibits the typical Bloch-oscillation dynamics. 

On the LPS, for an initial Fock state $|j\rangle$, the Husimi distribution is $Q(|\beta|)=|J_l(2|\beta|)|^2$, independent of $\mathrm{arg}\,\beta$ and hence of $k$, so the Fock-state Husimi function encircles the cylinder. During evolution, $Q(\beta,t)=Q(|\beta|,t)$ remains independent of $k$, while its width along the cylinder oscillates with frequency $\omega$. If, instead, the initial state is localized in $k$ with a small width $\Delta k$, then $Q(\beta)$ is stretched along the cylinder (width $\sim 1/\Delta k$), and during evolution it both moves up and down the cylinder and encircles it simultaneously.

For $\omega=0$, equation~(\ref{e2momentum}) tells us that the spectrum forms a band $\varepsilon(k)=2J\cos(k)$ with the quasi-momentum $k\in(-\pi,+\pi]$, and the eigenstates are the extended Bloch states~\cite{sakurai2020modern}.

%%%%%%%%%%%%%%%%%%%%%%%%%%%%%%%%%%%%%%%%%%%%%%%%%%%%%%%%%%%%%%%%%%%%%%%%%%%%%%%%%%%%%%%%%%%%%%%%%%%%%%%%%%

\subsubsection{Heisenberg--Weyl algebra}\label{ssec:hw}
The Heisenberg--Weyl algebra $\mathfrak{hw}$ shares certain similarities with the Euclidean algebra, but differs in several important aspects. Mathematically, $\mathfrak{hw}$ is spanned by the generators $\{\hat{a},\hat{a}^\dagger,\mathbb{I}\}$, where $\mathbb{I}$ denotes the identity operator. Since $\mathbb{I}$ commutes with all other elements of the algebra, it does not belong to the Cartan subalgebra but instead generates a central extension of the algebra. In physical applications it is often convenient to enlarge the set of generators to $\{\hat{a},\hat{a}^\dagger,\hat{n},\mathbb{I}\}$, where the number operator $\hat{n}=\hat{a}^\dagger\hat{a}$ may be viewed as a Cartan generator.

In contrast to the $\mathfrak{e}(2)$ algebra, the root generators of $\mathfrak{hw}$ include normalization factors,
\begin{equation}\label{nfactor}
    \hat{a}|n\rangle = \sqrt{n}\,|n-1\rangle, \qquad
    \hat{a}^\dagger|n\rangle = \sqrt{n+1}\,|n+1\rangle,
\end{equation}
and the existence of a vacuum state $|0\rangle$ imposes a lower boundary on the FSL, which therefore admits a natural reference state $|\psi_0\rangle=|0\rangle$. The nonvanishing commutation relation
\begin{equation}
    [\hat{a},\hat{a}^\dagger] = \mathbb{I}
\end{equation}
implies the absence of a Hermitian phase operator $\hat{\phi}$ canonically conjugate to $\hat{n}$, i.e.,
$[\hat{n},\hat{\phi}] = i\mathbb{I}$, unless the Hilbert space is truncated to finite dimension~\cite{susskind1964quantum,pegg1989phase}.

Despite this, coherent states $|\alpha\rangle$, defined via the displacement operator
\begin{equation}\label{bosondisplace}
    \hat{D}_\mathrm{b}(\alpha)
    = e^{\alpha\hat{a}^\dagger - \alpha^*\hat{a}}, \qquad
    |\alpha\rangle = \hat{D}_\mathrm{b}(\alpha)|0\rangle,
\end{equation}
are well behaved. They are Gaussian minimum-uncertainty states and form an overcomplete basis, with overlap
\begin{equation}
    |\langle\beta|\alpha\rangle|^2
    = e^{-|\alpha-\beta|^2}.
\end{equation}
In this case, the Lie-algebra phase space coincides with the regular phase space of a bosonic mode or a particle on a line, namely the $x$--$p$ plane. Unlike the $\mathfrak{e}(2)$ algebra, the displacement operators do not commute, but instead satisfy
\begin{equation}
    \hat{D}_\mathrm{b}(\alpha)\hat{D}_\mathrm{b}(\beta)
    = e^{i\,\mathrm{Im}(\alpha\beta^*)}\hat{D}_\mathrm{b}(\alpha+\beta).
\end{equation}

The Cartan generator $\hat{n}$ is diagonal in the bosonic Fock basis $|n\rangle$, while the root generators $\hat{a}$ and $\hat{a}^\dagger$ connect neighboring Fock states in the FSL. Consequently, the FSL is an infinite one-dimensional tight-binding chain with sites labeled by $n=0,1,\dots$. The normalization factors in Eq.~(\ref{nfactor}) imply that the tunneling amplitudes are site dependent, scaling as $\sqrt{n}$.

To draw an analogy between the LPS and the FSL similar to Fig.~\ref{fig:e2lps} for the $\mathfrak{e}(2)$ algebra, we note that the LPS is the $x$--$p$ plane, while the Cartan generator $\hat{n}$ may be viewed as defining an axis perpendicular to this plane. Introducing the parabolic surface
\begin{equation}
    n(x,p) = \frac{p^2}{2} + \frac{x^2}{2} + \frac{1}{2},
\end{equation}
one may relate the Husimi function $Q(x,p)$ to the Fock-state distribution by first projecting $Q(x,p)$ onto this surface and subsequently projecting onto the $n$ axis.

As a physical realization, consider the driven harmonic oscillator,
\begin{equation}
    \hat{H}_\mathrm{dHO}
    = \Delta\hat{n} + \eta\left(\hat{a}+\hat{a}^\dagger\right),
\end{equation}
where $\Delta$ is the detuning between the oscillator and the drive, and $\eta$ is the drive amplitude~\cite{scully1997quantum,saugmann2023fock}. Since the Heisenberg--Weyl algebra is finite dimensional as a Lie algebra, the Heisenberg equations of motion close and the system is analytically solvable. Here it is clear that as long as $\Delta\neq0$ we need the extended algebra, $\{\hat{a},\hat{a}^\dagger,\hat{n},\mathbb{I}\}$ instead of $\{\hat{a},\hat{a}^\dagger,\mathbb{I}\}$. We note that the Hamiltonian can alternatively be diagonalized using the displacement $\hat{D}_\mathrm{b}(-\eta/\Delta)$, yielding eigenenergies $\varepsilon_n=\Delta n-\eta^2/\Delta$ and eigenstates given by displaced Fock states.

%%%%%%%%%%%%%%%%%%%%%%%%%%%%%%%%%%%%%%%%%%%%%%%%%%%%%%%%%%%%%%%%%%%%%%%%%%%%%%%%%%%%%%%%%%%%%%

\subsubsection{$\mathfrak{su}(2)$ algebra}
A spin-$S$ particle has a finite Hilbert space with Fock states $|S,m\rangle$ and a compact phase space forming a coherent-state sphere of radius $S$. For the regular phase space, we consider distributions such as the Husimi function, living on this sphere. In the LPS, coherent states correspond to points on the sphere, often referred to as the \textit{Bloch sphere}. Extending this to distributions, the LPS can be identified with the standard phase-space description for a spin-$S$ particle.

The $\mathfrak{su}(2)$ algebra is given by the standard angular momentum commutation relations
\begin{equation}
    [\hat{S}^+,\hat{S}^-] = 2\hat{S}_z, \qquad
    [\hat{S}_z,\hat{S}^\pm] = \pm \hat{S}^\pm,
\end{equation}
with different values of $S$ corresponding to different representations. The Casimir operator $\hat{\mathbf{S}}^2$ sets the conserved total spin value.

We take the reference state (vacuum) to be $|\psi_0\rangle = |S,S\rangle$, localized at the north pole. Introducing polar and azimuthal angles $\theta\in[0,\pi]$ and $\phi\in[0,2\pi)$, we define the complex displacement parameter
\begin{equation}
    \varsigma = -\frac{\theta}{2} e^{i\phi},
\end{equation}
and the spin displacement operator~\cite{zhang1990coherent}
\begin{equation}\label{spindisplace}
    \hat{D}(\varsigma) = e^{\varsigma \hat{S}^+ - \varsigma^* \hat{S}^-}.
\end{equation}
The resulting Perelomov coherent states (which coincide with standard spin-coherent states) are
\begin{equation}
    |\theta,\phi\rangle = \hat{D}(\varsigma) |S,S\rangle.
\end{equation}
If $\mathbf{n}$ denotes the unit vector along the coherent-state direction and $\hat{S}_\perp$ is the spin component perpendicular to $\mathbf{n}$, then
\begin{equation}
    (\Delta \hat{S}_\perp)^2 = \frac{\hbar^2 S}{2}.
\end{equation}
The spin-coherent states form an overcomplete basis,
\begin{equation}
    \frac{2S+1}{4\pi} \int \sin\theta\, d\theta\, d\phi\, 
    |\theta,\phi\rangle \langle \theta,\phi| = \mathbb{I}.
\end{equation}
In the spin Fock basis, they expand as
\begin{equation}\label{spinexp}
\begin{aligned}
    |\theta,\phi\rangle
    &= \sum_{m=-S}^{S} \binom{2S}{S+m}^{1/2}
    \left(\cos\frac{\theta}{2}\right)^{S+m}
    \left(\sin\frac{\theta}{2}\right)^{S-m} \\
    &\quad \times e^{-i(S-m)\phi}\, |S,m\rangle.
\end{aligned}
\end{equation}

Choosing $\hat{S}_z$ as the Cartan generator, the FSL population can be visualized as the projection of the phase-space distribution onto the $z$-axis of the Bloch sphere. The corresponding FSL distribution is
\begin{equation}
    P(m) = |\langle S,m|\theta,\phi\rangle|^2,
\end{equation}
with variance
\begin{equation}
    \Delta m = \sqrt{\frac{S}{2}}\,|\sin\theta|,
\end{equation}
which vanishes at the poles and reaches a maximum at the equator ($\theta = \pi/2$). This reflects the broken translational invariance of the FSL and highlights that localization in LPS does not automatically imply localization in the FSL.

For a more flexible FSL representation, a semisimple algebra like $\mathfrak{su}(2)$ can be realized using two bosonic modes via the \textit{Schwinger-boson mapping}~\cite{sakurai2020modern}:
\begin{equation}
    \hat{S}_z = \frac{1}{2}\left(\hat{a}^\dagger\hat{a}-\hat{b}^\dagger\hat{b}\right),\quad
    \hat{S}^+ = \hat{a}^\dagger\hat{b},\quad
    \hat{S}^- = \hat{b}^\dagger\hat{a}.
\end{equation}
The Casimir operator maps to the total boson number,
\begin{equation}
    \hat{N} = \hat{a}^\dagger\hat{a} + \hat{b}^\dagger\hat{b},
\end{equation}
and the spin Fock states $|S,m\rangle$ correspond to bosonic number states $|n_a,n_b\rangle$ with $n_a+n_b = 2S$.

\begin{figure}[!ht]
    \centering
    \includegraphics[width=1\linewidth]{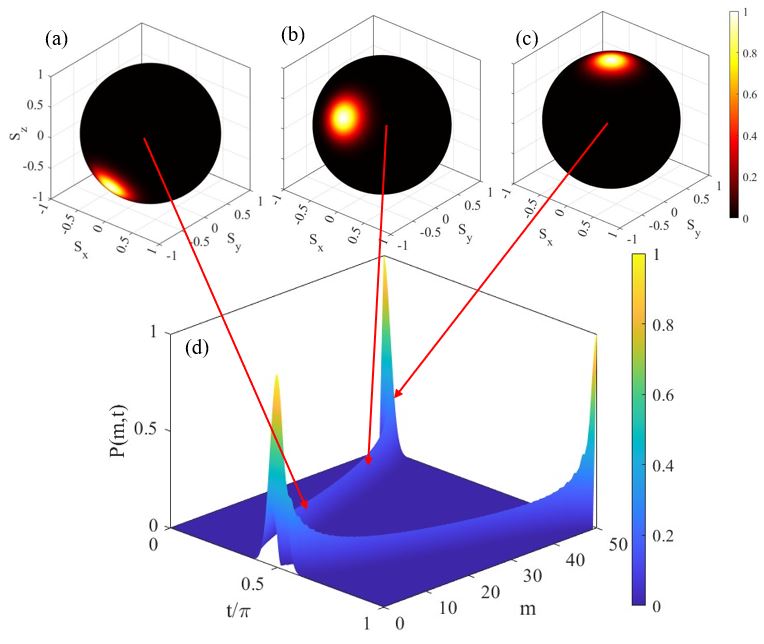}
    \caption{Demonstration of the time evolution in the FSL (d) and the LPS (a)–(c) for the $\mathfrak{su}(2)$ model. Panel (d) shows the time evolution of the FSL population $P(m,t)$, while panels (a)–(c) display the Husimi function $Q(\theta,\phi)$ on the Bloch sphere. The initial state is $|\psi(0)\rangle = |S,S\rangle$ with $S=50$, corresponding to a Perelomov coherent state, and the dynamics are governed by the Hamiltonian~(\ref{su2ham}) with tunneling rate $J=1$. In the FSL, the state starts at one edge of the lattice and undergoes oscillatory dynamics with revival time $T_\mathrm{rev}=\pi$, such that at half the revival time the population is completely transferred to the opposite edge. On the LPS, since the Hamiltonian generates a displacement, the state remains coherent throughout the evolution and follows a geodesic (great circle) on the Bloch sphere. Panels (a), (b), and (c) correspond to times $t=0$, $t=\pi/6$, and $t=\pi/3$, respectively, with red arrows indicating the corresponding FSL distributions.
}
    \label{fig:su2ev}
\end{figure}

Considering the two-mode Hamiltonian $\hat{H}_\mathrm{2m}$ of Eq.~(\ref{su2ham}) as a realization of the $\mathfrak{su}(2)$ Fock-state lattice (FSL), the system exhibits perfect state transfer between the two modes~\cite{christandl2004perfect,saugmann2023fock}. For an initial state $|\psi_a\rangle \otimes |0\rangle$, the state is perfectly transferred—up to phase factors—to mode $b$ after half the revival time $T_\mathrm{rev}/2$, reflecting the equidistant spectrum of the FSL. This behavior is illustrated in Fig.~\ref{fig:su2ev} for the initial Fock state $|N,0\rangle$. In addition to the time evolution of the FSL population $P(m,t)$, the figure also shows three snapshots of the corresponding Husimi function, which follows a geodesic on the Bloch sphere.

Let us end with a comment on the Lie-algebra solution~(\ref{lieheis}) of the Heisenberg equations. Consider the Hamiltonian 
\[
\hat{H}_{\mathfrak{su}(2)} = \delta \hat{S}_z + 2J \hat{S}_x
\] 
or its two-mode boson representation 
\[
\hat{H}_{\mathfrak{su}(2)} = \frac{\delta}{2} (\hat{n}_a - \hat{n}_b) + J (\hat{a}^\dagger \hat{b} + \hat{b}^\dagger \hat{a}).
\] 
We can solve for either the spin operators or the boson operators (e.g., via a Bogoliubov diagonalization). The first yields solutions like $\hat{S}_x(t)$ or equivalently $(\hat{a}^\dagger \hat{b} + \hat{b}^\dagger \hat{a})(t)$, while the latter gives $\hat{a}(t)$ and $\hat{b}(t)$. Specifically, assuming the initial state $|S,S\rangle$, the analytical solutions are
\[
\begin{aligned}
\langle \hat{S}_x(t) \rangle &= \frac{2J \delta}{\Omega^2} \, S \, \bigl[ 1 - \cos(\Omega t) \bigr],\\
\langle \hat{S}_y(t) \rangle &= - \frac{2J}{\Omega} \, S \, \sin(\Omega t),\\
\langle \hat{S}_z(t) \rangle &= S \, \frac{\delta^2 + (2J)^2 \cos(\Omega t)}{\Omega^2},\\
\langle \hat{a}(t) \rangle &= \sqrt{2S} \left( \cos\frac{\Omega t}{2} - i \frac{\delta}{\Omega} \, \sin\frac{\Omega t}{2} \right),\\
\langle \hat{b}(t) \rangle &= - i \sqrt{2S} \, \frac{2J}{\Omega} \, \sin\frac{\Omega t}{2},
\end{aligned}
\]
with $\Omega = \sqrt{\delta^2 + (2J)^2}$. 

Since the boson operators $\hat{a}$ and $\hat{b}$ are not elements of the Lie algebra, the last two equations contain information not captured by the Lie-algebra solution (the first three equations). In other words, while the Heisenberg equations~(\ref{lieheis}) for the $\mathfrak{su}(2)$ generators close and yield exact dynamics for all observables within the algebra, the Bogoliubov approach provides a finer-grained description at the level of the underlying modes. The Lie-algebraic solution captures the collective degrees of freedom, whereas the Bogoliubov method resolves the microscopic mode structure.

%%%%%%%%%%%%%%%%%%%%%%%%%%%%%%%%%%%%%%%%%%%%%%%%%%%%%%%%%%%%%%%%%%%%%%%%%%%%%%%%%%%%%%%%%%%%%%%%%%%%%%%%%%%%%%%%%%%%%%%%%%%

\subsubsection{$\mathfrak{su}(3)$ algebra}
The $\mathfrak{su}(3)$ algebra differs from the previous examples in that it possesses two Cartan generators, resulting in a two-dimensional FSL. In addition, there are three root pairs, giving rise to six root generators, where each pair defines a displacement operator $\hat{D}_{\alpha_i}(\beta_i)$ [Eq.~(\ref{gendis})]. Consequently, each lattice site in the FSL has six nearest neighbors, forming a triangular lattice which was already mentioned in subsec.~\ref{ssec:fsl}. 

For $\mathfrak{su}(2)$ the minimal matrix representation is given by the Pauli matrices, while for $\mathfrak{su}(3)$ it is the $3\times3$ Gell-Mann matrices $\hat{\lambda}_k$ ($k=1,2,\dots,8$). Analogous to the Pauli matrices, the Gell-Mann matrices can be generalized to larger representations. More relevant for FSLs, as in $\mathfrak{su}(2)$, a Schwinger-boson mapping allows the Cartan and root generators to be expressed in terms of bosonic modes~\cite{anishetty2009irreducible}, as introduced in Eq.~(\ref{su3ham}). The operators $\hat{H}_{1}$ and $\hat{H}_{2}$ act as the Cartan generators and are diagonal in the Fock basis. The remaining operators, $\hat{I}_{\pm}$, $\hat{U}_{\pm}$, and $\hat{V}_{\pm}$, are root generators, generally denoted $\hat{E}_{\alpha}$ and labeled by roots $\alpha$. In the lattice plane, not all three roots are independent; rather, $\alpha_3=\alpha_1+\alpha_2$, consistent with the three primitive directions of a triangular lattice. As in $\mathfrak{su}(2)$, the sum of squared generators defines the quadratic Casimir operator $\hat{\Gamma}_2$, while a second, cubic Casimir operator $\hat{\Gamma}_3$ involves products of three generators. The operator $\hat{\Gamma}_2$ enforces conservation of the total boson number, so the Fock states can be written as
\begin{equation}\label{su3fock}
    |n_a,n_b,n_c\rangle = |n_a,n_b,N-n_a-n_b\rangle,
\end{equation}
with $N=n_a+n_b+n_c$ the total boson number. The physical interpretation of $\hat{\Gamma}_3$ is less direct, but it is associated with a chirality in the FSL connecting the three modes.

In the previous examples, we could visualize the mapping from a two-dimensional LPS to a one-dimensional FSL. Here, the LPS is four dimensional while the FSL is two dimensional, rendering such visualization impractical. Instead, we focus on understanding the FSL structure through the bosonic representation~\cite{saugmann2023fock}. For fixed $N$, the triangular FSL contains $(N+1)(N+2)/2$ sites corresponding to the Fock states in Eq.~(\ref{su3fock}). The three corners correspond to single-mode occupation, e.g., $|N,0,0\rangle$, while sites along an edge have exactly one empty mode, e.g., $|n_a,N-n_a,0\rangle$. More generally, along any primitive lattice direction one bosonic mode remains fixed, and the corresponding root-generator pair, e.g., $(\hat{I}_{-},\hat{I}_{+})$, induces nearest-neighbor hopping along that direction. Accordingly, the displacement operator
\begin{equation}
    \hat{D}_{I}(\beta_I) = e^{\beta_I \hat{I}_{+} - \beta_I^* \hat{I}_{-}},
\end{equation}
and analogously for the $\hat{U}$ and $\hat{V}$ generators, generates translations along the lattice directions associated with $\alpha_1$, $\alpha_2$, and $\alpha_3$, respectively.

Any of the corner states may be chosen as the reference state $|\psi_0\rangle$, and all $\mathfrak{su}(3)$ PCSs can be generated from it using the operator
\begin{equation}
    \hat{\Omega}(\xi)
    = e^{\xi_1 \hat{a}^\dagger\hat{b} + \xi_2 \hat{b}^\dagger\hat{c} + \xi_3 \hat{a}^\dagger\hat{c} - \mathrm{H.c.}}, \qquad \xi_i \in \mathbb{C},
\end{equation}
such that $|\xi\rangle = \hat{\Omega}(\xi)|N,0,0\rangle$. Because the displacement operators $\hat{D}_{\alpha_i}(\beta_i)$ do not mutually commute, $\hat{\Omega}(\xi)$ cannot be factorized into a simple product of displacement operators, reflecting the noncommutativity of the algebra (rotations in higher dimensions).

An alternative representation of the $\mathfrak{su}(3)$ PCSs is given by~\cite{perelomov1977generalized,Perelomov1986,mathur2001coherent}
\begin{equation}
    |\xi\rangle = \frac{1}{\sqrt{N!}}
    \left( \zeta_1 \hat{a}^\dagger + \zeta_2 \hat{b}^\dagger + \zeta_3 \hat{c}^\dagger \right)^N
    |0,0,0\rangle.
\end{equation}
For the four-dimensional LPS, the coefficients $\zeta_i$ may be parametrized by four angles as
\begin{equation}
    \begin{aligned}
         \zeta_1 &= \cos\theta_1, \\
         \zeta_2 &= e^{i\phi_1} \sin\theta_1 \cos\theta_2,\\
         \zeta_3 &= e^{i\phi_2} \sin\theta_1 \sin\theta_2,
    \end{aligned}
\end{equation}
with $0 \le \theta_1,\theta_2 \le \pi/2$ and $0 \le \phi_1,\phi_2 \le 2\pi$. In principle, one can also derive an explicit mapping between the parameter sets $\{\zeta_i\}$ and $\{\xi_i\}$.

\begin{figure}[!ht]
    \centering
    \includegraphics[width=0.6\linewidth]{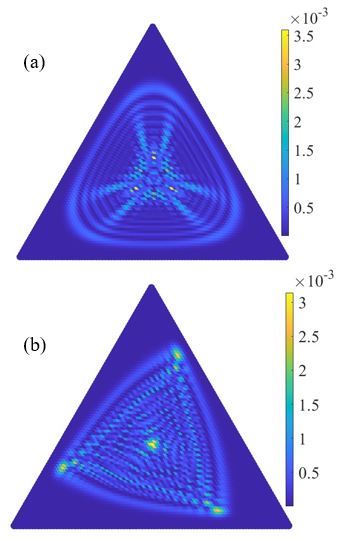}
    \caption{Two snapshots of the distribution $P(n_a,n_b,N-n_a-n_b,t)$ under the $\mathfrak{su}(3)$ Hamiltonian~(\ref{su3ham2}). 
        The initial state is $|30,30,30\rangle$, i.e., $n_a=n_b=n_c=30$, corresponding to the center site of the lattice. 
        Panel (a) shows the time-reversal symmetric case ($\phi=0$), while panel (b) has $\phi=\pi/3$, introducing a synthetic magnetic flux in the FSL and breaking the symmetry seen in (a). 
        Both snapshots are taken at $t=0.3J$ with total boson number $N=90$.}
    \label{fig:su3ev}
\end{figure}

We now turn to the evolution in the FSL under the Hamiltonian
\begin{equation}\label{su3ham2}
    \hat{H}_{\mathfrak{su}(3)}
    = J\left(\hat{a}^\dagger\hat{b}
    + \hat{b}^\dagger\hat{c}
    + e^{i\phi}\hat{a}^\dagger\hat{c}
    + \mathrm{H.c.}\right).
\end{equation}
The phase factor $e^{i\phi}$ introduces a staggered synthetic magnetic flux through the triangular FSL, breaking time-reversal symmetry and inducing chiral motion, consistent with a nonvanishing local topological marker. The dynamics of this model, including chiral edge currents, have been studied in detail in Ref.~\cite{saugmann2023fock}. 

Figure~\ref{fig:su3ev} shows two snapshots of the distribution $P(n_a,n_b,N-n_a-n_b,t)$ evolved under Hamiltonian~(\ref{su3ham2}). In both panels, the initial state is $|30,30,30\rangle$, corresponding to the central lattice site. In panel (a), the dynamics are time-reversal symmetric ($\phi=0$), while in panel (b) time-reversal symmetry is broken by choosing $\phi=\pi/3$, resulting in a synthetic magnetic field in the FSL. The presence of a gauge field clearly breaks the symmetry observed in panel (a). In both cases, $t=0.3J$ and the total boson number is $N=90$.

%%%%%%%%%%%%%%%%%%%%%%%%%%%%%%%%%%%%%%%%%%%%%%%%%%%%%%%%%%%%%%%%%%%%%%%%%%%%%%%%%%%%%%%%%%%%%%%%%%%%%%%%%%%%%%%%%%%%%%%%%%%%%

\subsubsection{$\mathfrak{so}(5)$ algebra}

To the best of our knowledge, the FSL emerging from the $\mathfrak{so}(5)$ algebra has not been discussed previously. As it turns out, it exhibits several interesting properties, some of which we discuss here. The minimal representation of the $\mathfrak{so}(5)$ algebra consists of $5\times5$ real antisymmetric matrices, ten in total. The generators are associated with rotations in the different coordinate planes, analogous to how the generators of $\mathfrak{su}(2)$ and $\mathfrak{su}(3)$ realize complex rotations. Like $\mathfrak{su}(3)$, the $\mathfrak{so}(5)$ algebra has two Cartan generators, $\hat{H}_{1,2}$, resulting in a two-dimensional FSL. There are four pairs of root generators $\hat{\Sigma}_{\pm\alpha}$, which can be divided into \emph{short} roots ($\alpha_1,\,\alpha_2$) and \emph{long} roots ($\alpha_1+\alpha_2,\,\alpha_1-\alpha_2$). These generate tunneling between nearest- and next-nearest-neighbor sites in the FSL, respectively. Consequently, the FSL forms a square lattice with both horizontal/vertical and diagonal tunneling terms. Since the LPS is six-dimensional and compact, the corresponding FSL is also compact. In addition to the quadratic Casimir operator $\hat{\Gamma}_2$ (boson conservation), the algebra admits a quartic Casimir operator $\hat{\Gamma}_4$, which involves products of four generators.

The bosonic representation of the $\mathfrak{so}(5)$ algebra involves four modes,
$\hat{a}_{\uparrow,\downarrow}$ and $\hat{b}_{\uparrow,\downarrow}$. One might naively expect that two bosonic modes, $\hat{a}$ and $\hat{b}$, would suffice to construct a square FSL with both nearest- and next-nearest-neighbor tunneling. However, such a construction would necessarily involve generators that do not conserve the total boson number, such as $\hat{a}^\dagger\hat{b}^\dagger$ and $\hat{a}\hat{b}$, rendering the corresponding FSL noncompact. This situation instead corresponds to the algebra $\mathfrak{su}(1,1)\oplus\mathfrak{su}(1,1)$.

In the bosonic realization of $\mathfrak{so}(5)$, the Cartan generators may be chosen as
\begin{equation}
    \begin{aligned}
         \hat{H}_1 &= \frac{1}{2}\left(\hat{a}_\uparrow^\dagger\hat{a}_\uparrow
         - \hat{a}_\downarrow^\dagger\hat{a}_\downarrow\right), \\
         \hat{H}_2 &= \frac{1}{2}\left(\hat{b}_\uparrow^\dagger\hat{b}_\uparrow
         - \hat{b}_\downarrow^\dagger\hat{b}_\downarrow\right),
    \end{aligned}
\end{equation}
with root generators
\begin{equation}
    \begin{aligned}
        \hat{\Sigma}_{\alpha_1} &= \hat{a}_\uparrow^\dagger\hat{a}_\downarrow, \qquad
        \hat{\Sigma}_{\alpha_2} = \hat{b}_\uparrow^\dagger\hat{b}_\downarrow, \\
        \hat{\Sigma}_{\alpha_1+\alpha_2} &= \hat{a}_\uparrow^\dagger\hat{b}_\downarrow, \qquad
        \hat{\Sigma}_{\alpha_1-\alpha_2} = \hat{a}_\downarrow^\dagger\hat{b}_\uparrow,
    \end{aligned}
\end{equation}
together with their Hermitian conjugates corresponding to the negative roots. This construction, involving four bosonic modes divided into two $a$ and two $b$ modes, is standard: the labels $a$ and $b$ denote orbital degrees of freedom, while the subscripts indicate an internal spin.

The Fock states are of the form
$|n_{a_\uparrow},n_{a_\downarrow},n_{b_\uparrow},n_{b_\downarrow}\rangle$,
with the total boson number
$N=n_{a_\uparrow}+n_{a_\downarrow}+n_{b_\uparrow}+n_{b_\downarrow}$
being conserved. The lattice-site indices $(i,j)$ of the FSL are determined by the
eigenvalues of the Cartan generators,
$m_1=n_{a_\uparrow}-n_{a_\downarrow}$ and
$m_2=n_{b_\uparrow}-n_{b_\downarrow}$.
Fixing the total boson number restricts the lattice to a finite square with a diamond-shaped boundary in the $(i,j)$ plane. The corner sites of this diamond correspond to extremal Fock states such as $|N,0,0,0\rangle$, $|0,N,0,0\rangle$.

It is instructive to note that
$\left(\hat{H}_1,\hat{\Sigma}_{\pm\alpha_1}\right)$ and
$\left(\hat{H}_2,\hat{\Sigma}_{\pm\alpha_2}\right)$
each form an $\mathfrak{su}(2)$ subalgebra. The remaining root generators, which correspond to diagonal tunneling processes in the FSL, are essential for closing the full $\mathfrak{so}(5)$ algebra. Their presence also implies that the FSL cannot be decomposed into independent lower-dimensional lattices. Algebraically, this reflects the fact that
$\mathfrak{so}(5) \not\simeq \mathfrak{su}(2) \oplus \mathfrak{su}(2)$.

To discuss FSL dynamics, we consider the Hamiltonian
\begin{equation}\label{so5ham}
\begin{aligned}
\hat H_{\mathfrak{so}(5)} &= 
J_1 \Big(\hat a_\uparrow^\dagger \hat a_\downarrow
+ \hat b_\uparrow^\dagger \hat b_\downarrow
+ \text{H.c.} \Big) \\
&+ J_2 \Big(
e^{i \phi} \, \hat a_\uparrow^\dagger \hat b_\uparrow
+ \hat a_\uparrow^\dagger \hat b_\downarrow
+ \hat a_\downarrow^\dagger \hat b_\uparrow
+ \hat a_\downarrow^\dagger \hat b_\downarrow
+ \text{H.c.} \Big),
\end{aligned}
\end{equation}
where $J_1$ and $J_2$ are the tunneling amplitudes for nearest- and next-nearest-neighbor processes, respectively, and the phase factor $\phi$ generates a synthetic flux in the FSL. Although several closed loops can be drawn within a square plaquette containing vertical, horizontal, and diagonal bonds, these loops are not independent. The square loop decomposes into sums of triangular loops, and gauge freedom removes all but a single gauge-invariant phase. Consequently, the dynamics is governed by a single independent flux parameter, even though some closed paths accumulate zero net phase.

For fixed $N$, the Hilbert-space dimension is
$\mathrm{Dim}(\mathcal{H}) = N^2/2 + N + 1$,
which equals the number of sites in the FSL. Since $\hat{H}_{\mathfrak{so}(5)}$ is quadratic, it can be diagonalized by first diagonalizing the corresponding single-particle Hamiltonian
\begin{equation}\label{so5ham2}
\hat{h}_{\mathfrak{so}(5)} = 
\begin{pmatrix}
0 & J_1 & J_1 & J_2 e^{i \phi} \\
J_1 & 0 & J_2 e^{i \phi} & J_1 \\
J_1 & J_2 e^{-i \phi} & 0 & J_1 \\
J_2 e^{-i \phi} & J_1 & J_1 & 0
\end{pmatrix}.
\end{equation}
The eigenvalues are roots of a quartic polynomial and are generally cumbersome. However, for $J_1 = J_2 = J$ one finds
\begin{equation}
\epsilon_{1,2,3,4} = \pm J\sqrt{2 \pm 2\cos\left(\frac{\phi}{2}\right)} ,
\end{equation}
with the full many-body spectrum given by
\begin{equation}
E_{\{n_i\}} = \sum_{i=1}^{4} \epsilon_i\, n_i,
\qquad 
n_i \in \{0,1,\dots,N\},
\end{equation}
subject to the constraint $\sum_{i=1}^{4} n_i = N$. If
\begin{equation}
    \frac{\epsilon_1}{\epsilon_3}=\sqrt{\frac{2+2\cos\left(\phi/2\right)}{2-2\cos\left(\phi/2\right)}}=\cot\left(\frac{\phi}{4}\right)
\end{equation}
is a rational number $m/n$, the system exhibits perfect revivals after a time
$T_\mathrm{rev} = \pi m/[J\cos(\phi/4)]$. This revival behavior does not persist for generic values of $J_1$ and $J_2$.

\begin{figure}[!ht]
    \centering
    \includegraphics[width=1\linewidth]{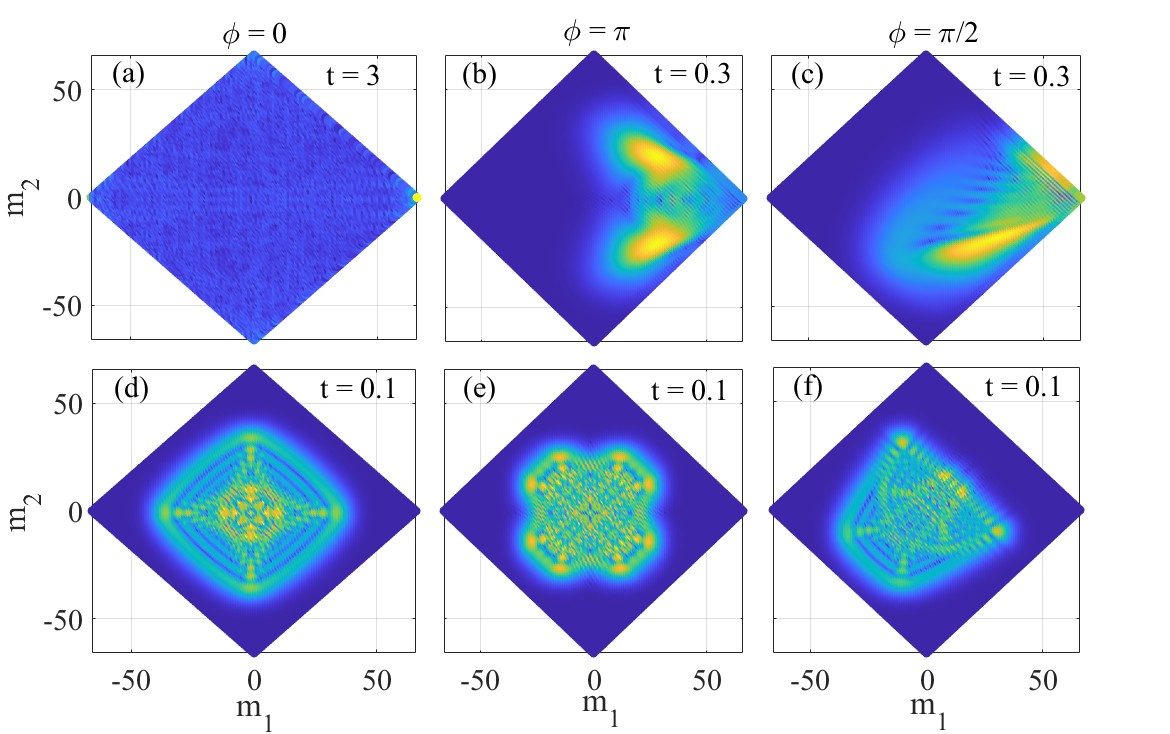}
    \caption{Snapshots of the fourth root of the distribution $P(m_1,m_2,t)$ in the FSL corresponding to the $\mathfrak{so}(5)$ model, evolved under the Hamiltonian~(\ref{so5ham}). The upper row corresponds to the initial state $|n_{a_\uparrow}=132,n_{a_\downarrow}=0,n_{b_\uparrow}=0,n_{b_\downarrow}=0\rangle$,while the lower row shows evolution from the central lattice site $|n_{a_\uparrow}=33,n_{a_\downarrow}=33,n_{b_\uparrow}=33,n_{b_\downarrow}=33\rangle$. The phase is $\phi=0$ in panels (a) and (d), $\phi=\pi$ in panels (b) and (e), and $\phi=\pi/2$ in panels (c) and (f). For $\phi=\pi/2$, time-reversal symmetry is broken and the distribution exhibits asymmetric chiral evolution. All plots are shown for the balanced case $J_1=J_2=1$ with total boson number $N=132$, and the time $t$ for each example is given in the panels.}
    \label{fig:so5ev}
\end{figure}

Examples of the FSL distribution for various initial states and values of $\phi$ are shown in Fig.~\ref{fig:so5ev}. For visual clarity, we plot the fourth root of the distribution. For $\phi=0$ and $\phi=\pi$, the Hamiltonians~(\ref{so5ham}) and~(\ref{so5ham2}) are real and preserve time-reversal symmetry, leading to symmetric evolution. For $\phi=\pi/2$, by contrast, the distribution preferentially flows in one direction, giving rise to a chiral current in the lattice.

%\onecolumngrid
%\twocolumngrid

Let us end by noting an additional peculiarity of the $\mathfrak{so}(5)$ FSL. 
The general rule regarding the dimensionality of Hamiltonian FSLs is that for every additional boson degree of freedom, the dimension of the FSL increases by one. 
A spin-$S$ provides a quasi-dimension, i.e., finite in length. 
On the other hand, a symmetry splits the lattice into corresponding sectors for the conserved quantity. 
Effectively, a continuous symmetry reduces the dimensionality by one, and hence~\cite{saugmann2023fock}
\begin{equation}\label{dim}
\text{Dim(FSL)} \;=\; \text{\# boson modes} \;-\; \text{\# contin. symmetries}.
\end{equation}

Although the boson representation of $\mathfrak{so}(5)$ involves four modes and the FSL is two-dimensional, this does not imply the existence of two continuous symmetries. 
The single $U(1)$ symmetry (boson number conservation) fixes the irreducible representation, while the two-dimensional lattice arises from the rank-2 Cartan structure of $\mathfrak{so}(5)$. 
As pointed out earlier, the additional Casimir operators determine the FSL geometry and label the representation, but do not generate independent symmetries or conserved directions. 
Thus, while Eq.~(\ref{dim}) is almost always correct, the $\mathfrak{so}(5)$ FSL provides an example where it does not hold; more generally, in the case where there exists an underlying Lie algebra we have
\begin{equation}
\text{Dim(FSL)} = \text{rank}(\mathfrak{g}).
\end{equation}

%%%%%%%%%%%%%%%%%%%%%%%%%%%%%%%%%%%%%%%%%%%%%%%%%%%%%%%%%%%%%%%%%%%%%%%%%%%%%%%%%%%%%%%%%%%%%%%%%%

\subsubsection{$\mathfrak{su}(1,1)$ algebra}
Like for the $\mathfrak{so}(5)$ algebra, the FSL of the $\mathfrak{su}(1,1)$ one does not seem to be discussed in the past. As we shall see below, it is related to Hamiltonians describing various types of field squeezing. It is conceptually different from say $\mathfrak{e}(2)$, $\mathfrak{hw}$, and $\mathfrak{su}(2)$ where there is a mapping between local states in the LPS and local states in the FSL. Here, instead, locality in LPS does not automatically imply locality in the FSL.

The $\mathfrak{su}(1,1)$ algebra has commutation relations similar to $\mathfrak{su}(2)$ up to a sign change,
\begin{equation}
    [\hat{K}^+,\hat{K}^-]=-2\hat{K}_0, \quad [\hat{K}_0,\hat{K}^\pm]=\pm \hat{K}^\pm.
\end{equation}
This seemingly minor change has profound consequences: the algebra becomes noncompact. The Casimir operator 
\begin{equation}\label{su11cas}
    \hat{\Gamma}=\hat{K}_0^2-\hat{K}_1^2-\hat{K}_2^2,
\end{equation}
where $\hat{K}^\pm = \hat{K}_1 \pm i \hat{K}_2$. The form~(\ref{su11cas}) implies that the associated LPS has a hyperboloid geometry. The Bergmann index $k$, related to $\hat{\Gamma}$, depends on the chosen representation. Notably, $\mathfrak{su}(1,1)$ is isomorphic to the Lorentz algebra $\mathfrak{so}(2,1)$, reflecting the form of the Casimir operator. The larger $\mathfrak{so}(3,1)$ is also noncompact, but has two Cartan generators such that its FSL becomes a 2D square lattice.

With a single Cartan generator $\hat{K}_0$, the corresponding FSL is one-dimensional, with Fock states $|k,m\rangle$ where $m$ is a non-negative integer. A standard boson representation is
\begin{equation}\label{smsu11}
    \hat{K}_0 = \frac{1}{2}\left(\hat{a}^\dagger \hat{a} + \frac{1}{2}\right), \quad
    \hat{K}^+ = \frac{1}{2} (\hat{a}^\dagger)^2, \quad
    \hat{K}^- = \frac{1}{2} \hat{a}^2.
\end{equation}
Both the vacuum $|0\rangle$ and the single excitation $|1\rangle$ are annihilated by $\hat{K}^-$, allowing either to serve as a reference state, $|\psi_0^{(0,1)}\rangle$. Choosing $|0\rangle$ yields $k=1/4$, while $|1\rangle$ gives $k=3/4$, corresponding to even- and odd-parity FSLs, respectively. The action of the generators on these states is~\cite{buvzek19901}:
\begin{equation}
\begin{aligned}
    \hat{K}_0 |k,m\rangle &= (k+m)|k,m\rangle,\\
    \hat{K}^+ |k,m\rangle &= \sqrt{(m+1)(m+2k)}\,|k,m+1\rangle,\\
    \hat{K}^- |k,m\rangle &= \sqrt{m(m+2k-1)}\,|k,m-1\rangle.
\end{aligned}
\end{equation}

The displacement operator
\begin{equation}
    \hat{D}(\beta) = e^{\beta (\hat{a}^\dagger)^2 - \beta^* \hat{a}^2}
\end{equation}
is the familiar single-mode squeezing operator $\hat{S}(\xi)$ with $\xi = \beta/2$. The corresponding PCS is $|k,\xi\rangle$, dependent on the reference state. The expectations of the generators, $K_0=\langle k,\xi|\hat{K}_0|k,\xi\rangle$ and $K^\pm=\langle k,\xi|\hat{K}^\pm|k,\xi\rangle$, are~\cite{duan2023quantum}
\begin{equation}\label{kexp}
    \left(K_0,K^+,K^-\right)=k\left(\cosh2r,\,\sinh2r\cos\theta,\,\sinh2r\sin\theta\right),
\end{equation}
where $\xi=re^{i\theta}$. Moreover, for the vacuum ($k=1/4$), it expands in the Fock basis as~\cite{mandel1995optical}
\begin{equation}\label{sqvac}
    |\xi\rangle \equiv \hat{S}(\xi)|0\rangle
    = \frac{1}{\sqrt{\cosh r}} \sum_{n=0}^{\infty} \frac{\sqrt{(2n)!}}{2^n n!} \left(-e^{i\theta} \tanh r\right)^n |2n\rangle,
\end{equation}
from where we can extract the boson distribution $p(n,\xi)=|\langle n|\xi\rangle|^2$. We also find the quadrature variances
\begin{equation}
\begin{array}{l}
    (\Delta x)^2 = \frac{1}{2}(\cosh 2r - \cos\theta \sinh 2r), \\ \\
    (\Delta p)^2 = \frac{1}{2}(\cosh 2r + \cos\theta \sinh 2r).
    \end{array}
\end{equation}

Let us comment on the origin of the two distinct representations, $k=1/4$ and $k=3/4$, that arise when using the single-mode bosonic realization~(\ref{smsu11}). The associated FSL is the set of non-negative integers, but the choice of reference state—either $|0\rangle$ or $|1\rangle$—restricts the dynamics to even or odd Fock states, respectively. In the Hamiltonian language, this is reflected in parity symmetry, $\hat{a}\rightarrow-\hat{a}$ and $\hat{a}^\dagger\rightarrow-\hat{a}^\dagger$, leaving the generators~(\ref{smsu11}) invariant. Thus, the FSL is actually two 1D chains, and similarly the LPS will also be two hyperboloid surfaces, one for each $k$.

Now suppose the state evolves under the Hamiltonian $\hat{H}=\omega \hat{a}^\dagger \hat{a}$ for a time $T=2\pi/\omega$. The quadrature operators $\hat{x}(t)$ and $\hat{p}(t)$ then undergo a full $2\pi$ rotation in phase space, and accordingly the Husimi distribution $Q(\alpha)$ completes one full revolution. The generalized Husimi distribution in the LPS, however, performs two circuits around the hyperboloid. This reflects the fact that the generators $\hat{K}^\pm$ are quadratic in the bosonic operators and therefore return to themselves already after $T=\pi/\omega$. The necessity of a doubled LPS thus follows directly from the quadratic structure of the algebra. As a side note, upon completing a full $2\pi$ rotation, the state acquires a geometric phase associated with the curvature of the LPS manifold~\cite{chiao1988lorentz,hong1988berry}. In the FSL, such a geometric contribution will not, however, manifest in the distribution $p(n,\xi)$.

An additional subtlety arises when attempting to relate the FSL and LPS. For both the $\mathfrak{hw}$ and $\mathfrak{su}(2)$ algebras, the mapping can be visualized as perpendicular projections once the appropriate surfaces are identified. This construction relies on the specific parametrization of those phase spaces. In the present case, however, the LPS is a hyperboloid while the FSL corresponds to the $\hat{K}_0$ (``$z$'') axis. The parametrization $\xi = r e^{i\phi}$, together with the form of the Casimir operator~(\ref{su11cas}), implies that the projection from the LPS onto the FSL is not perpendicular.

This is evident when considering the squeezed vacuum: its FSL distribution $p(n,\xi)$ corresponding to eq.~(\ref{sqvac}) is peaked at $n=0$ and decays monotonically with increasing $n$, whereas the generalized Husimi distribution~(\ref{genhusim}) is localized at a finite distance from the bottom of the hyperboloid. Consequently, the intuitive geometric correspondence between localization in the LPS and population along the FSL differs qualitatively from the cases discussed earlier. Since the mapping from the LPS to the FSL is nonlinear and shaped by the curvature of the underlying group manifold, localization properties are not preserved under projection. Thus, states that are well localized in the LPS may correspond to broad distributions on the FSL.

As for the other cases, let us consider the evolution within the FSL under a Hamiltonian
\begin{equation}
    \hat{H}_{\mathfrak{su}(1,1)}=\omega\hat{a}^\dagger\hat{a}+\frac{1}{2}\left(\xi\left(\hat{a}^\dagger\right)^2+\xi^*\hat{a}^2\right).
\end{equation}
As initial state we consider the vacuum, such that the evolved state is a squeezed vacuum as in equation~(\ref{sqvac}) but with time-dependent parameters and an overall phase factor. More precisely, the evolved state will be 
\begin{equation}
    |\psi(t)\rangle=e^{i\chi(t)}|\zeta(t)\rangle,
\end{equation}
where
\begin{equation}
    \chi=\frac{\omega t}{2}+\frac{1}{2}\arctan\left(\frac{\omega}{\Omega}\tan(\Omega t)\right)
\end{equation}
and if we parametrize $\zeta(t)=r(t)e^{i\theta(t)}$, we have~\cite{mandel1995optical}
\begin{equation}
    \begin{array}{cll}
        \tanh r(t) & = & \displaystyle{\frac{|\xi|}{\sqrt{\omega^2+\Omega^2\cot^2(\Omega t)}}},\\ \\
        \theta(t) & = & \displaystyle{\theta - \frac{\pi}{2} - \arctan\Big( \frac{\omega}{\Omega} \, \tan(\Omega t) \Big)},
    \end{array}
\end{equation}
with $\Omega=\sqrt{\omega^2-|\xi|^2}$. It is clear that there are two types of solutions, whether $\omega^2-|\xi|^2$ is positive or negative, leading to stable oscillatory solutions or unstable exponential runaway solutions. The average boson number 
\begin{equation}
    \langle\hat{n}\rangle_t=\frac{|\xi|^2}{\Omega^2}\sin^2(\Omega t),
\end{equation}
with a variance $\text{Var}(n)_t = 2  \langle \hat{n} \rangle_t \left(\langle \hat{n} \rangle_t + 1\right)$. 

For completeness, we note two other bosonic realizations of $\mathfrak{su}(1,1)$. First, the single-mode representation~\cite{buck1981exactly}:
\begin{equation}
    \hat{K}_0 = \hat{a}^\dagger \hat{a} + \frac{1}{2}, \quad
    \hat{K}^+ = \sqrt{\hat{a}^\dagger \hat{a}} \, \hat{a}^\dagger, \quad
    \hat{K}^- = \hat{a} \sqrt{\hat{a}^\dagger \hat{a}},
\end{equation}
with $k=1/2$, relevant for Jaynes-Cummings models with intensity-dependent couplings~\cite{buvzek1989jaynes}, and related to more general Holstein-Primakoff transformations~\cite{holstein1940field}. Second, the two-mode squeezing representation~\cite{gerry1991correlated,gilles1992non}:
\begin{equation}
    \hat{K}_0 = \frac{1}{2}(\hat{a}^\dagger \hat{a} + \hat{b}^\dagger \hat{b} + 1), \quad
    \hat{K}^+ = \hat{a}^\dagger \hat{b}^\dagger, \quad
    \hat{K}^- = \hat{a} \hat{b},
\end{equation}
with $k = (|n_a - n_b| + 1)/2$. Also for these representations, the corresponding FSL distributions $p(n,\xi)$ will exponentially decay for increasing $n$'s, and with widths $\Delta n$ growing exponentially with $|\xi|$.

%%%%%%%%%%%%%%%%%%%%%%%%%%%%%%%%%%%%%%%%%%%%%%%%%%%%%%%%%%%%%%%%%%%%%%%%%%%%%%%%%%%%%

%\onecolumngrid

%\twocolumngrid
%%%%%%%%%%%%%%%%%%%%%%%%%%%%%%%%%%%%%%%%%%%%%%%%%%%%%%%%%%%%%%%%%%%%%%%%%%%%%%%%%%%%%
\section{Discussions and further aspects}
\label{sec:discussion}
We have seen using several examples how a Lie algebra $\mathfrak{g}$ defines a FSL by assigning Fock states to the sites of the weight lattice, and using the root generators to define the non-vanishing tunneling rates. Moreover, from the elements $\hat{X}_a$ of $\mathfrak{g}$ we can form a Hamiltonian via a linear combination as in~(\ref{lieham}). In this case, the Lie-algebra and the Hamiltonian FSLs are identical up to factors multiplying the tunneling rates, i.e., assuming all $k_a=1$ in~(\ref{lieham}) reproduces the Lie algebra FSL. Thus, starting from a Lie algebra we constructed its FSL and a family of Hamiltonians. 

Historically, FSLs have been discussed where the Hamiltonian alone defines the lattice, without reference to an underlying Lie algebra. Yet, if such an algebra exists, it can provide further insight, for example by identifying a LPS or simplifying analytical derivations. We therefore ask whether there is a reverse connection: given $\hat{H}$, does there exist an underlying Lie algebra? It is clear that this is a strong statement: it is not enough for the Hamiltonian to be symmetric under a Lie group; their FSLs must coincide.  
We focus on finite algebras, i.e., the Hamiltonian has a finite number of degrees of freedom. If the Hamiltonian is non-integrable, we typically have no closed set of Heisenberg equations, and we cannot expect a finite underlying algebra. Hence, we consider integrable cases, the simplest being 
\begin{equation}
    \hat{H}_{2b}=\sum_{i,j=1}^N\left(A_{ij}\hat{a}_i^\dagger\hat{a}_j
    +\frac{1}{2}B_{ij}\hat{a}_i^\dagger\hat{a}_j^\dagger
    +\frac{1}{2}B_{ij}^*\hat{a}_i\hat{a}_j\right),
\end{equation}
for bosonic modes $\hat{a}_i$ with $A$ Hermitian and $B$ symmetric. The elements of the underlying Lie algebra are
\begin{equation}
    \left\{\hat{a}_i^\dagger\hat{a}_j,\hat{a}_i^\dagger\hat{a}_j^\dagger,\hat{a}_i\hat{a}_j\right\}_{i,j=1}^N.
\end{equation}
It is straightforward to show that this forms a closed set under commutation; this is the bosonic representation of the \textit{symplectic algebra} $\mathfrak{sp}(2N)$. The Cartan generators are $\hat{a}_i^\dagger\hat{a}_i$, and the root generators can be grouped into hopping terms $\hat{a}_i^\dagger\hat{a}_j$ ($i\neq j$), pair creation $\hat{a}_i^\dagger\hat{a}_j^\dagger$, and pair annihilation $\hat{a}_i\hat{a}_j$. The FSL is then a hypercube, possibly with tunneling beyond nearest neighbors. Adding symmetries reduce the set of elements and we can find subalgebras of $\mathfrak{sp}(2N)$, like many of those discussed in the previous section. 

Adding pump terms linear in creation/annihilation operators does not break integrability. From subsection~\ref{ssec:hw}, the set $\{\hat{a},\hat{a}^\dagger,\hat{\mathbb{I}}\}$ forms the (non-extended) $\mathfrak{hw}$ algebra. Indeed, the set
\begin{equation}
    \left\{\hat{a}_i,\hat{a}_i^\dagger,\hat{a}_i^\dagger\hat{a}_j,
    \hat{a}_i^\dagger\hat{a}_j^\dagger,\hat{a}_i\hat{a}_j,\hat{\mathbb{I}}\right\}_{i,j=1}^N
\end{equation}
forms a closed algebra called the inhomogeneous symplectic Lie algebra $\mathfrak{isp}(2N)$,
\begin{equation}
    \mathfrak{isp}(2N) = \mathfrak{sp}(2N) \ltimes \mathfrak{hw}_N.
\end{equation}
Here, $\ltimes$ denotes the \textit{semidirect product}, meaning that the $N$-mode Heisenberg-Weyl algebra $\mathfrak{hw}_N$ is a \textit{normal ideal} of $\mathfrak{isp}(2N)$; that is, $[g,h]\in \mathfrak{hw}_N$ if $h\in \mathfrak{hw}_N$ and $g\in \mathfrak{sp}(2N)$. Since the algebra contains the central element $\hat{\mathbb{I}}$, it is not semisimple. 

For fermions $\hat{c}_i$ and $\hat{c}_i^\dagger$ with antisymmetric $B$, the corresponding algebra
\begin{equation}
    \left\{\hat{c}_i^\dagger\hat{c}_j,\hat{c}_i^\dagger\hat{c}_j^\dagger,\hat{c}_i\hat{c}_j\right\}_{i,j=1}^N
\end{equation}
is $\mathfrak{so}(2N)$, and the FSL is a finite hypercube $\{0,1\}^N$. Extending as in the bosonic case,
\begin{equation}
    \left\{\hat{c}_i,\hat{c}_i^\dagger,\hat{c}_i^\dagger\hat{c}_j,
    \hat{c}_i^\dagger\hat{c}_j^\dagger,\hat{c}_i\hat{c}_j,\hat{\mathbb{I}}\right\}_{i,j=1}^N,
\end{equation}
also forms a closed algebra, though without a standard name. %In terms of Majorana fermions,
%\begin{equation}
%    \hat{\gamma}_{2i-1}=\hat{c}_i+\hat{c}_i^\dagger, \quad
%    \hat{\gamma}_{2i}=i(\hat{c}_i-\hat{c}_i^\dagger),
%\end{equation}
%which satisfy $\hat{\gamma}_i=\hat{\gamma}_i^\dagger$, the elements
%\begin{equation}
%    \left\{\hat{\gamma}_i,\hat{\gamma}_i\hat{\gamma}_j\right\}_{i,j=1}^{2N}
%\end{equation}
%span $\mathfrak{so}(2N+1)$, and we do not need to include the identity $\hat{\mathbb{I}}$ to get a closed algebra. 

We now consider examples where different degrees of freedom combine. For instance, the Jaynes-Cummings model~\cite{larson2021jaynes},
\begin{equation}
    \hat{H}_\mathrm{JC}=\omega \hat{a}^\dagger\hat{a}+\frac{\Omega}{2}\hat{\sigma}_z
    + g\left(\hat{a}^\dagger \hat{\sigma}^- + \hat{\sigma}^+ \hat{a}\right),
\end{equation}
has Pauli operators $\hat{\sigma}$. Due to particle conservation, the FSL is an infinite set of two-site lattices, one for each conserved excitation number. Although integrable, the Heisenberg equations do not close, so no finite-dimensional underlying Lie algebra exists. Constructing an infinite-dimensional algebra from all commutators would yield an infinite rank and thus an infinite-dimensional FSL. Hence, integrability alone does not guarantee a finite Lie algebra. A similar argument applies to the Lipkin-Meshkov-Glick model~\cite{larson2021jaynes},
\begin{equation}\label{lmg}
    \hat{H}_\mathrm{LMG}=\Omega \hat{S}_z + \frac{g}{S} \hat{S}_x^2,
\end{equation}
which is solvable, but its Heisenberg equations do not close. Its FSL consists of two 1D finite chains, which cannot be reproduced from an infinite Lie algebra.

In quantum optics, models like Jaynes-Cummings combine bosonic and spin degrees of freedom. Using a Jordan-Wigner transformation~\cite{auerbach2012interacting}, a single spin-$1/2$ can be mapped to a fermion: $\hat{\sigma}_z=2\hat{c}^\dagger\hat{c}-1$, $\hat{\sigma}^+=\hat{c}^\dagger$, $\hat{\sigma}^-=\hat{c}$. This motivates \textit{Lie superalgebras}, where the algebra is $\mathfrak{g}=\mathfrak{g}_\mathrm{even}\oplus \mathfrak{g}_\mathrm{odd}$ with bosonic (even) and fermionic (odd) generators. The \textit{super-commutator} is defined as
\begin{equation}
    [\hat{x},\hat{y}\} =
    \begin{cases}
        [\hat{x},\hat{y}], & \text{$x$ and $y$ even},\\
        [\hat{x},\hat{y}], & x \text{ even},\, y \text{ odd},\\
        \{\hat{x},\hat{y}\}, & x \text{ odd},\, y \text{ odd}.
    \end{cases}
\end{equation}

For the Jaynes-Cummings model, the generators
\begin{equation}\label{jcalgebra}
    \{\hat{a}^\dagger \hat{a},\,\hat{c}^\dagger \hat{c},\,\hat{a}^\dagger \hat{c},\,\hat{c}^\dagger \hat{a}\},
\end{equation}
with Cartan generators $\hat{a}^\dagger\hat{a}$ and $\hat{c}^\dagger\hat{c}$ (even) and root generators $\hat{a}^\dagger \hat{c}$ and $\hat{c}^\dagger \hat{a}$ (odd), reproduce the FSL from $\hat{H}_\mathrm{JC}$. The corresponding Lie superalgebra is $\mathfrak{su}(1|1)$: one bosonic and one fermionic mode. If we try the same to the quantum Rabi model (Jaynes-Cummings including the counter-rotating terms $\hat{a}^\dagger\hat{\sigma}^+$ and $\hat{a}\hat{\sigma}^-$), we do not find a finite Lie superalgebra, despite the fact that the model is integrable~\cite{braak2011integrability}.

The `fermionization' can be generalized to the Tavis-Cummings model. The Jordan-Wigner transformation of spin operators $\hat{S}^\pm$ produces fermionic strings, so the transformed Hamiltonian $\hat{H}_\mathrm{TC}$ is no longer quadratic. Nevertheless, identifying Cartan generators $\hat{a}^\dagger\hat{a}$ and $\hat{S}_z$ (even) and root generators $\hat{a}^\dagger \hat{S}^-$ and $\hat{a} \hat{S}^+$ (odd), we see the superalgebra $\mathfrak{su}(1|N)$, with $N=2S$ fermionic modes. The corresponding fermionic Hilbert space dimension is $2^N$, typically much larger than $2S+1$, as the Jordan-Wigner mapping involves all spin sectors. Also here, the Hamiltonian and Lie superalgebra FSLs agree.

Extending to superalgebras thus allows identification of underlying algebras for a larger set of integrable Hamiltonians. However, like the quantum Rabi model, mapping the Lipkin-Meshkov-Glick Hamiltonian~(\ref{lmg}) to a fermionic model does not yield a finite Lie (super)algebra, so the inversion procedure is not generally possible.

%%%%%%%%%%%%%%%%%%%%%%%%%%%%%%%%%%%%%%%%%%%%%%%%%%%%%%%%%%%%%%%%%%%%%%%%%%%%%%%%%%%%%

\section{Conclusion}
\label{sec:conclusion}
In this work, rather than analyzing FSLs solely as structures induced by specific Hamiltonians, we have taken an alternative approach by introducing them from the perspective of Lie algebras. More precisely, a Lie algebra defines a FSL through its generators: the eigenvalues of the Cartan generators label the sites of the weight lattice, and by identifying these sites with eigenvectors (Fock states) and introducing tunneling processes governed by the root generators, one obtains a \emph{Lie-algebra FSL}. Once such a FSL is defined, it can naturally be associated with a phase space intrinsic to the Lie algebra. In particular, phase-space distributions such as the Husimi function $Q(\beta)$ live on the manifold of generalized coherent states. This Lie phase space (LPS) can differ significantly from the phase spaces commonly encountered in textbooks on quantum optics. For example, while the $\mathfrak{su}(1,1)$ algebra (relevant for squeezing dynamics) can be represented using a single bosonic mode and described in the usual flat $x$--$p$ phase space, it may alternatively be represented on its LPS, which is a curved two-dimensional hyperboloid.

When an FSL admits an underlying Lie algebra structure, the dynamics can therefore be analyzed equivalently in the LPS. As in standard phase-space formulations of quantum mechanics, this representation can offer additional physical insight by encoding quantum states in terms of real-valued distributions. In particular, the curvature of the LPS can give rise to nontrivial geometric effects, such as geometric phases familiar from spin systems. It can be shown~\cite{zhang1990coherent,gazeau2009coherent,arnol2013mathematical} that when the Hamiltonian is linear in the generators, the Husimi function~(\ref{genhusim}) evolves according to
\begin{equation}
    \partial_t Q = \{Q,H_\mathrm{cl}\}_\mathrm{LP},
\end{equation}
where $H_\mathrm{cl}$ is the corresponding classical Hamiltonian and the \emph{Lie--Poisson bracket} is defined as
\begin{equation}
    \{F,G\}_\mathrm{LP} = {f_{ab}}^{c}\, x_c
    \frac{\partial F}{\partial x_a}
    \frac{\partial G}{\partial x_b},
\end{equation}
with $x_a=\langle\hat{X}_a\rangle$ and structure constants ${f_{ab}}^{c}$ identical to those in Eq.~(\ref{liecom}). This formulation generalizes the familiar phase-space evolution equations encountered in conventional quantum mechanics.

Beyond establishing a link between FSLs and phase space, we have highlighted several key insights. While the dimensionality of a FSL is often dictated by the number of bosonic modes and continuous symmetries, an underlying Lie-algebraic structure can lead to important exceptions, as demonstrated by the $\mathfrak{so}(5)$ example. Furthermore, the relation between localization in the LPS and localization in the FSL is generally nontrivial, particularly for noncompact algebras such as $\mathfrak{su}(1,1)$, where curvature induces nonlinear projections between the two descriptions.

We have also examined the inverse problem: starting from an integrable Hamiltonian, constructing its FSL, and asking whether a Lie algebra exists that reproduces the same lattice structure. For quadratic bosonic and fermionic Hamiltonians, such underlying algebras can indeed be identified. However, for integrable models combining different types of degrees of freedom—such as the Jaynes--Cummings and Tavis--Cummings models—no finite Lie algebra exists. In these cases, Lie superalgebras provide a natural extension, allowing one to recover the correct FSL structure and offering an algebraic interpretation of the dynamics.

Importantly, Hamiltonian integrability alone does not guarantee the existence of an underlying Lie (super)algebra, as illustrated by the Lipkin--Meshkov--Glick model. Thus, while the algebraic construction of FSLs provides a powerful framework for understanding and visualizing quantum dynamics, it is not universally invertible starting from an arbitrary integrable Hamiltonian. Overall, our results establish a systematic foundation for exploring FSLs in both Lie-algebraic and superalgebraic settings, and open avenues for analytical and numerical investigations of quantum systems with structured Hilbert spaces.

%\bibliography{main}


\end{document}